\documentclass[aps,prb,twocolumn,superscriptaddress,showpacs]{revtex4-1}
\usepackage{graphicx,psfrag}
\usepackage{amsmath}

\begin{document}


\title{Magnetic order, magnetic correlations and spin dynamics in the pyrochlore antiferromagnet 
Er$_2$Ti$_2$O$_7$} 


\author{P.~Dalmas de R\'eotier}
\affiliation{Institut Nanosciences et Cryog\'enie, SPSMS, 
CEA and Universit\'e Joseph Fourier, F-38054 Grenoble, France}
\author {A.~Yaouanc}
\altaffiliation[Present address: ]{Laboratory for Muon-Spin Spectroscopy, 
Paul Scherrer Institute, 5232 Villigen-PSI, Switzerland}
\affiliation{Institut Nanosciences et Cryog\'enie, SPSMS, 
CEA and Universit\'e Joseph Fourier, F-38054 Grenoble, France}
\author {Y.~Chapuis}
\affiliation{Institut Nanosciences et Cryog\'enie, SPSMS, 
CEA and Universit\'e Joseph Fourier, F-38054 Grenoble, France}
\author{S.~H.~Curnoe}
\affiliation{Department of Physics and Physical Oceanography, Memorial University,
St.\ John's, NL, Canada}
\author {B.~Grenier}
\affiliation{Institut Nanosciences et Cryog\'enie, SPSMS, 
CEA and Universit\'e Joseph Fourier, F-38054 Grenoble, France}
\author {E.~Ressouche}
\affiliation{Institut Nanosciences et Cryog\'enie, SPSMS, 
CEA and Universit\'e Joseph Fourier, F-38054 Grenoble, France}
\author {C.~Marin}
\affiliation{Institut Nanosciences et Cryog\'enie, SPSMS, 
CEA and Universit\'e Joseph Fourier, F-38054 Grenoble, France}
\author{J.~Lago}
\affiliation{Department of Inorganic Chemistry, Universidade del Pa\'is Vasco, 48080 Bilbao, Spain}
\author{C.~Baines}
\affiliation{Laboratory for Muon-Spin Spectroscopy, 
Paul Scherrer Institute, 5232 Villigen-PSI, Switzerland}
\author{S.~R.~Giblin}
\affiliation{ISIS Facility, Rutherford Appleton Laboratory, Chilton, Didcot, OX11 0QX, UK}



\date{\today}

\begin{abstract}

Er$_2$Ti$_2$O$_7$ is believed to be a realization of an XY antiferromagnet on a frustrated lattice of corner-sharing regular tetrahedra. It is presented as an example of the order-by-disorder mechanism in which fluctuations lift the degeneracy of the ground state, leading to an ordered state. Here we report detailed measurements of the low temperature magnetic properties of Er$_2$Ti$_2$O$_7$, which displays a second-order phase transition at $T_{\rm N} \simeq 1.2$~K with coexisting short- and long-range orders. Magnetic-susceptibility studies show that there is no spin-glass-like irreversible effect. Heat-capacity measurements reveal that the paramagnetic critical exponent is typical of a 3-dimensional XY magnet while the low-temperature specific heat sets an upper limit on the possible spin-gap value and provides an estimate for the spin-wave velocity. Muon spin relaxation measurements show the presence of spin dynamics in the nanosecond time scale down to 21~mK. This time range is intermediate between the shorter time characterizing the spin dynamics in Tb$_2$Sn$_2$O$_7$, which also displays long- and short-range magnetic order, and the time scale typical of conventional magnets.  Hence the ground state is characterized by exotic spin dynamics. We determine the parameters of a symmetry-dictated Hamiltonian restricted to the spins in a tetrahedron, by fitting the paramagnetic diffuse neutron scattering intensity for two reciprocal lattice planes. These data are recorded in a temperature region where the assumption that the correlations are limited to nearest neighbors is fair.
\end{abstract}

\pacs{75.40.-s, 75.10.Dg, 76.75.+i}

\maketitle

\section{Introduction} 
\label{Introduction}

Because of the geometrical frustration of their magnetic
superexchange interactions, the insulating pyrochlore
compounds $R_2M_2$O$_7$, where $R$ stands for a magnetic
rare earth ion and $M$ = Ti or Sn, display a variety
of unusual magnetic behaviors.\cite{Gardner10} Examples include (i) spin
ice systems Ho$_2$Ti$_2$O$_7$ and Dy$_2$Ti$_2$O$_7$,\cite{Harris97,Ramirez99} 
(ii) Yb$_2$Ti$_2$O$_7$ with a sharp transition in the spin dynamics 
finger-printed by
a pronounced peak in the specific heat,\cite{Hodges02} and (iii) 
Tb$_2$Sn$_2$O$_7$ in which magnetic Bragg reflections are observed at low
temperature by neutron diffraction,\cite{Mirebeau05} while no spontaneous 
magnetic field is found by the muon spin rotation ($\mu$SR) technique.\cite{Dalmas06} In addition, even when a spontaneous 
field and magnetic Bragg reflections are detected, as it is expected for a conventional ordered magnet, 
persistent spin dynamics in the ordered state are surprisingly observed, e.g. in Gd$_2$Ti$_2$O$_7$ and 
Gd$_2$Sn$_2$O$_7$.\cite{Champion01,Stewart04,Wills06,Brammall11,Bertin02,Yaouanc05a,Chapuis09b}
In terms of crystal-field anisotropy, the spin-ice systems are strongly Ising-like. 
Tb$_2$Sn$_2$O$_7$ has also an Ising anisotropy, but not so strong. Yb$_2$Ti$_2$O$_7$ is XY-like 
from the crystal-field point of view and the Gd compounds are approximately isotropic. 

Although Yb$_2$Ti$_2$O$_7$ is XY-like, its magnetic moments are not perpendicular to the local ${<}111{>}$ axes and it does not display long-range magnetic order,\cite{Hodges01,Hodges02} although this absence of order has been disputed \cite{Yasui03,Gardner04} and is still under debate.\cite{Chang12} These anisotropy and absence of long-range order also pertain for Yb$_2$GaSbO$_7$.\cite{Hodges11} Hence, it was of great interest when Er$_2$Ti$_2$O$_7$ was reported to be XY-like and to display long-range order at low temperature with the Er$^{3+}$ magnetic moments perpendicular to their local $[ 111]$ axes.\cite{Champion03} Later on, however, coexisting short- and long-range orders were found and soft collective modes were detected.\cite{Ruff08} 
The presence of the soft modes has been  attributed to the 
incommensurate value of the canting angle.\cite{Briffa11} These astonishing 
inferences call for more detailed data and analysis. This is the purpose of this work. One of the experimental advantages of Er$_2$Ti$_2$O$_7$ over Yb$_2$Ti$_2$O$_7$ and Yb$_2$GaSbO$_7$, is the possibility to produce large high-quality crystals.

Another reason for the interest in the Er$_2$Ti$_2$O$_7$ system is the following. As a realization of an XY antiferromagnet on a pyrochlore lattice, it is a natural candidate for observing the phenomenon of order by disorder which has been discussed theoretically for more than three decades since the pioneering study by Villain and coworkers.\cite{Villain80} 
Bramwell {\em et al.} indeed showed that while the zero temperature ground state is degenerate, thermal fluctuations select a subset of the manifold and induce a first order phase transition to a conventional N\'eel ground state.\cite{Bramwell94} The order by 
disorder mechanism has been confirmed in several subsequent works, see e.g. Refs.~\onlinecite{Champion03,Champion04,Zhitomirsky12,Savary12a}, and interestingly the more recent studies consider the effect of quantum fluctuations and tend to explain the second order nature of the transition experimentally observed in Er$_2$Ti$_2$O$_7$.

The organization of this paper is as follows.  Section~\ref{Survey} gives a survey of the physical 
properties of Er$_2$Ti$_2$O$_7$ and discusses its magnetic structure. In Sec.~\ref{Experimental}
we describe the growth of the single crystals, their basic characterizations and the experimental methods
used in the present work. Section \ref{Results_bulk} presents our investigation of the bulk properties of the system, including magnetic susceptibility and specific heat measurements and their 
analysis. The following section (Sec.~\ref{Results_microscopic}) deals with the microscopic techniques, {\em i.e.} muon spin relaxation and neutron scattering in the paramagnetic phase.
A summary of our key results is given in Sec.~\ref{Summary}. 
The physics of effective one-half spins (Kramers doublets) on a tetrahedron is described in Appendix~\ref{Tetrahedron}. Appendix~\ref{Vzz} outlines the calculation needed for the analysis of the specific heat data presented in Sec.~\ref{Results_bulk}.

\section{Physical properties of E\lowercase{r}$_2$T\lowercase{i}$_2$O$_7$ and magnetic structure} 
\label{Survey}

Erbium titanate, Er$_2$Ti$_2$O$_7$, is an insulating pyrochlore compound that crystallizes
into the cubic space group $Fd{\bar 3}m$, with the lattice parameter $a = 10.0727\, (1)$~\AA\ at room 
temperature and $ x = 0.3278 \, (8)$, the free position parameter allowed by the space group
which characterizes the 48f site occupied by oxygen.\cite{Helean04}

\begin{figure}
\includegraphics[scale=0.75]{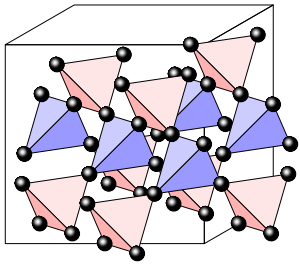}
\caption{(color online).
The network of corner sharing regular tetrahedra formed by the rare earth atoms in the pyrochlore structure in which Er$_2$Ti$_2$O$_7$ crystallizes.
The axis of trigonal symmetry at the position of a rare earth is one of the cube diagonals.
There are two types of tetrahedra in the network, which differ
by their orientation: type B is rotated by 90$^\circ$ about the cubic axes 
with respect to type A.  We distinguish the two sets by two colors in the drawing.
Since each rare earth is at a corner shared by two tetrahedra, one of each kind, 
either the set of the four corners of all the A tetrahedra or the set of the four corners
of all the B tetrahedra is sufficient to describe the Er$^{3+}$ lattice.
}
\label{pyrochlore_structure}
\end{figure}

The  Er$^{3+}$ ions which occupy the 16d Wyckoff positions in the space group,
are located at the vertices of a corner-sharing network of tetrahedra; 
see Fig.~\ref{pyrochlore_structure}. 
A single tetrahedron with four Er sites comprises the primitive unit cell.
The Er$^{3+}$ crystal sites are all equivalent and the local symmetry is 
$D_{3d}$, where the 3-fold axes pass through the center of a tetrahedron, in 
the directions $[111]$, $[\bar{1}\bar{1}1]$, $[\bar{1}1\bar{1}]$ and 
$[1\bar{1}\bar{1}]$ for the four sites numbered 1, 2, 3, and 4,
respectively on a single tetrahedron. According to Hund's rules, the total angular momentum of the Er$^{3+}$ ion 
in its ground multiplet is $J=15/2$. The 16-fold degeneracy 
is lifted into Kramers doublets by the crystal electric field (CEF).

The ground state doublet can be described as an effective spin $S=1/2$ with a
quantization axis $z$ parallel to the local trigonal axis. It is well isolated
from the excited doublets, the lowest being at about 74~K above the ground state in temperature units.\cite{Champion03}
We shall write the CEF ground state doublet wavefunctions as $| \phi_0^\pm \rangle$.
This doublet is characterized by its spectroscopic factors along and perpendicular 
to the trigonal axis, $g_\parallel$ and  $g_\perp$ respectively. From a global analysis
of the CEF for the pyrochlore $R_2$Ti$_2$O$_7$ series it has been deduced that
$g_\parallel = 2 g_J | \langle \phi_0^\pm | J_z | \phi_0^\pm \rangle | = 1.8 \, (5)$ and 
$g_\perp = g_J | \langle \phi_0^\pm | J_\pm | \phi_0^\mp \rangle | = 7.7 \, (1)$.\cite{Bertin12}
Here $g_J = 6/5$ is the Land\'e factor. 
These spectroscopic factors are related to matrix elements that we shall need for the analysis
of neutron scattering data. We have 
\begin{eqnarray}
j_{\rm CEF} & \equiv & \langle \phi_0^+ |J_z|\phi_0^+ \rangle =- \langle\phi_0^- |J_z|\phi_0^- \rangle 
= 0.75 \, (20), \\
t_{\rm CEF} & \equiv & \langle \phi_0^\pm |J_\pm|\phi_0^\mp\rangle = 6.42 \, (8).
\label{local_5}
\end{eqnarray}
where $J_{\pm} \equiv J_{x} \pm i J_{y}$.
The matrix elements obviously refer to quantities written in local axes; see 
Appendix~\ref{Tetrahedron_geometry} for a discussion. By definition,
the large difference between $j_{\rm CEF}$ and $t_{\rm CEF}$ (and obviously also  between $g_\parallel $ and $g_\perp $) 
reflects the strong CEF anisotropy of the XY type  of the Er spins  
(in contrast to the Ising limit for which $j_{\rm CEF}\gg t_{\rm CEF}$).

The compound displays a magnetic phase transition at $T_{\rm N} \simeq 1.2$~K.\cite{Blote69}
The large negative value of the Curie-Weiss temperature
$\theta_{\rm CW}$ ($\theta_{\rm CW}= -22$~K is deduced from susceptibility data measured between 20 and 50~K; see Refs.~\onlinecite{Blote69,Bramwell00})
suggests a strong antiferromagnet coupling. 

Neutron diffraction shows the magnetic structure below $T_{\rm N}$ to be noncollinear
with the propagation vector ${\bf k} = (0, 0, 0)$.\cite{Champion03}
From polarized neutron diffraction, the Er$^{3+}$ magnetic moment is determined
to be $m = 3.25 \, (9) \, \mu_{\rm B}$ at low temperature.\cite{Poole07}
A note of caution seems justified at this juncture: $m$ is not directly related to the 
spectroscopic factors which have been determined for a paramagnetic ion, since
the molecular field has to be taken into account for an estimation of $m$.
The diffraction data have been originally described \cite{Champion03,Poole07} with 
the $\Gamma_3^{+}$ irreducible representation.\footnote{In fact the authors of Refs. \onlinecite{Champion03} and \onlinecite{Poole07} label this representation as
$\Gamma_5$. Our labeling is
consistent with
http://www.cryst.ehu.es/cgi-bin/rep/programs/sam/point.py?sg=221}
We notice that this description also considered in Refs.~\onlinecite{Ruff08,Cao10,Sosin10} has been recently disputed by Briffa {\em et al.}\cite{Briffa11}

In fact the available microscopic information provides an insight into the moment orientation. 
Let us consider a one-half spin subjected to a molecular field 
oriented at a polar angle $\theta$ from the [111] axis. The following relation can be 
derived:\cite{Abragam70,Bonville78}
\begin{eqnarray}
\tan \theta = {g_\parallel \over g_\perp} 
\sqrt{{\mu^2_{\rm B} g^2_\parallel - 4 m^2 \over 4 m^2 - \mu^2_{\rm B} g^2_\perp}}.
\label{local_6}
\end{eqnarray}
Numerically this gives $\theta \approx 20^\circ$. This means that the field is not far from 
being parallel to the [111] axis. However, the polar angle $\phi$ of ${\bf m}$ is given by 
$\tan \phi = (g_\perp/g_\parallel)^2\tan \theta$, i.e. $\phi \approx 80^\circ$. Taking into account the uncertainties 
on $g_\perp$, $g_\parallel$, and $m$, this analysis indicates that the moment is 
perpendicular to the local [111] axis, or at least close to being perpendicular. This is consistent with the
magnetic structure first proposed by Champion {\it et al.},\cite{Champion03} and invalidates the proposal of Ref.~\onlinecite{Briffa11}.

\begin{figure}
\includegraphics[width=0.4\textwidth]{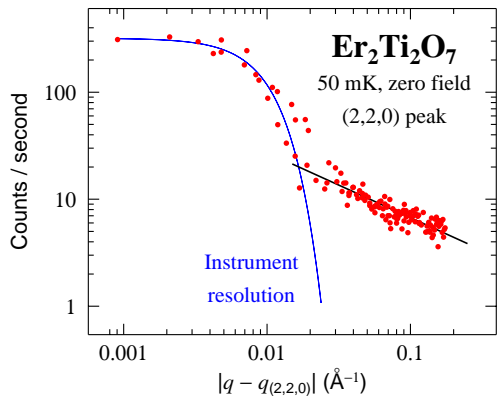}
\caption{(color online) 
Neutron intensity for the (2,2,0) Bragg peak of Er$_2$Ti$_2$O$_7$ plotted as a function of 
$|q - q_{(2,2,0)}|$. The data are from Ruff {\it et al.}\cite{Ruff08} The instrumental
resolution  is determined from the reflection measured under
a 3~T magnetic field. The line for $|q - q_{(2,2,0)}| \geq 0.015$~\AA$^{-1}$ is described in the main text. 
}
\label{Er2Ti2O7_PRL_101_147205_fig2f_reworked}
\end{figure}

Regardless of the magnetic structure, the (2,2,0) Bragg reflection has been shown to be  anomalous 
with broad tails.\cite{Ruff08} Quantitatively, we find that the intensity around this position is proportional to $|q - q_{(2,2,0)}|^{-\eta}$ with $\eta \simeq 0.6$ for 
$|q - q_{(2,2,0)}| > 0.03$~\AA$^{-1}$; 
see Fig.~\ref{Er2Ti2O7_PRL_101_147205_fig2f_reworked}.
The value of the exponent $\eta$ is much reduced compared to $\eta \simeq 1.8$ found in Tb$_2$Sn$_2$O$_7$.\cite{Dalmas06,Chapuis07} For a 
conventional ordered
compound the neutron intensity would be Gaussian-like, i.e. with a much steeper slope.
	
Inelastic neutron scattering data recorded in zero field suggest the presence of magnetic soft 
modes,\cite{Ruff08} in agreement with the power law behavior of the specific 
heat below $T_{\rm N}$.\cite{Sosin10} At first sight this is surprising given the expected strong crystal-field
anisotropy of the Er$^{3+}$ ions.\cite{Briffa11}

\section{Experimental} 
\label{Experimental}

\begin{figure}
\includegraphics[width=0.4\textwidth]{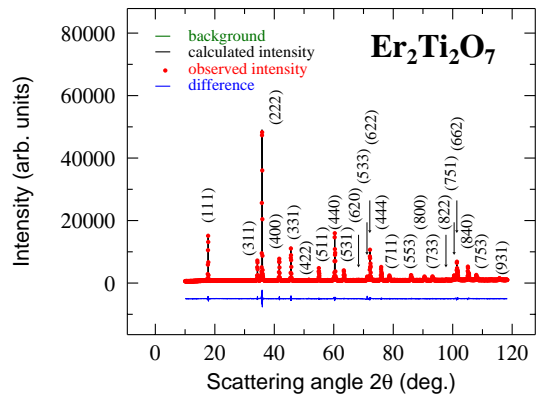}
\caption{(color online).
An example of an x-ray diffraction pattern recorded at room temperature for an Er$_2$Ti$_2$O$_7$ powder obtained after crushing part of a crystal. The radiation used is Co K$_\alpha$. The line at the bottom of the graph shows the difference between the data and the refinement model. }
\label{Experimental_xray}
\end{figure}

\begin{figure}
\includegraphics[width=0.4\textwidth]{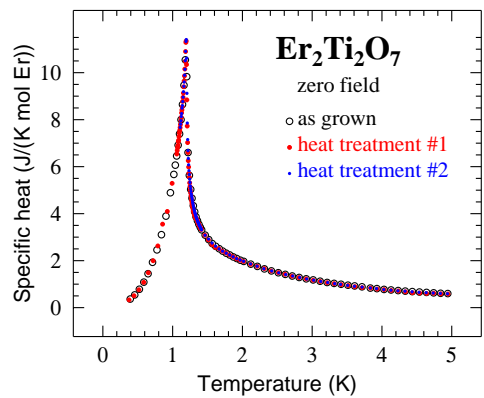}
\caption{(color online).
Specific heat versus temperature for three samples cut from an Er$_2$Ti$_2$O$_7$ crystal prepared as indicated in the main text. For the first sample, no subsequent treatment has been performed, while the other two result from post-growth heat treatments as follows. Case \#1: heat treatment for 7 days under oxygen at 1150$^\circ$C, followed by a slow cooling down 
to 400$^\circ$C. Case \#2: heat treatment for 7 days under argon at 1150$^\circ$C,
followed by a slow cooling down to 400$^\circ$C.
}
\label{Experimental_Cp}
\end{figure}

Er$_2$Ti$_2$O$_7$ single crystals were grown by the floating
zone technique using a commercial optical furnace.
Feed rods were prepared from high purity oxides (TiO$_2$,
99.995\% and Er$_2$O$_3$, 99.99\%), mixed and heat
treated up to 1180$^\circ$C. After sintering, a rod was heat
treated up to 1350$^\circ$C in air. Crystal growth conditions
were optimized under air (1 $\ell$/mn) at the growth rate
of 2~mm/h coupled with a rotation rate of 30~rounds per minute.
As with most of the titanate pyrochlores, Er$_2$Ti$_2$O$_7$ crystallized
rods are mostly transparent, with a slight pink color. No 
phases other than the cubic one with $Fd\bar{3}m$ space group were detected by
x-ray powder diffraction experiments; see Fig.~\ref{Experimental_xray} 
for an example. As grown and post growth heat-treated crystals were
characterized by specific heat measurements. No noticeable
differences were detected; see Fig.~\ref{Experimental_Cp}. 
This is in contrast to the case of Tb$_2$Ti$_2$O$_7$.\cite{Yaouanc11a}
The position of the peak provides a measure of the critical temperature.
We get 
$T_{\rm N} = 1.23 \, (1)$~K.
For comparison the following values have already been published:
$1.25$~K (Ref.~\onlinecite{Blote69}) and $1.173 \, (2)$~K (Ref.~\onlinecite{Champion03}) in reasonable agreement.

The investigation of the macroscopic properties of the system consisted of 
magnetic susceptibility experiments performed with a commercial magnetometer (Magnetic Property Measurement System, Quantum Design inc.) down to 2~K, and of 
heat capacity measurements. For this latter physical property, the temperature range from 0.48 to 20~K was investigated with a commercial calorimeter (Physical Property Measurement System, (PPMS) Quantum Design inc.) equipped with a $^3$He stage using a standard thermal
relaxation method. Additional measurements between 0.11 and 2.50~K were performed
with a home-made calorimeter inserted in a dilution refrigerator
using a semi-adiabatic technique.

The $\mu$SR measurements were carried out at
the European Muon Spectrometer of the ISIS facility
(Rutherford Appleton Laboratory, United Kingdom) and the 
Low Temperature Facility of the Swiss Muon Source (S$\mu$S, Paul Scherrer Institute,
Switzerland). The muon beam is pulsed at the former
facility and pseudo-continuous at the latter.

The neutron scattering experiments were performed at the Institut Laue 
Langevin (ILL, Grenoble) with the lifting-counter diffractometer D23 of the CEA collaborating Research Group (CRG). 

\section{Bulk measurements} 
\label{Results_bulk}

Here we shall first discuss magnetic susceptibility measurements.
Then we shall present zero-field specific heat results and finish with the 
determination of the magnetic phase diagram using specific heat-data recorded under magnetic
fields.
 
\subsection{Magnetic susceptibility} 
\label{Results_bulk_sus}

Since the magnetization measurements were performed on a needle-shape sample and the field 
was applied along its long axis, the demagnetization field is negligible.

Classically, the static magnetic susceptibility $\chi$ is expected to follow a Curie-Weiss law 
far from the ordering temperature in the paramagnetic regime. It reads
\begin{eqnarray}
\chi = {C \over T - \theta_{\rm CW}},
\label{Results_bulk_sus_Weiss_1}
\end{eqnarray}
where the Curie constant $C$ can be expressed in terms of the so-called paramagnetic 
moment $m_{\rm para}$:
\begin{eqnarray}
C = {1 \over v} {\mu_0 \, m^2_{\rm para} \over 3 k_{\rm B}}, 
\label{Results_bulk_sus_Weiss_2}
\end{eqnarray}
where $v = a^3/ N_{\rm cell}$ with $N_{\rm cell}$ being the number of Er$^{3 +}$ ions in the cubic 
cell ($N_{\rm cell}=16$). For an isolated Er$^{3 +}$ ion,  
$m_{\rm para} = g_J \sqrt {J (J +1)} \, \mu_{\rm B} = 9.58 \, \mu_{\rm B}$. 

\begin{figure}
\includegraphics[width=0.4\textwidth]{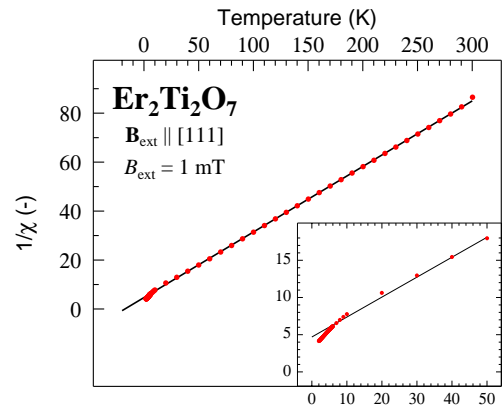}
\caption{(color online) 
Inverse of the magnetic susceptibility of an Er$_2$Ti$_2$O$_7$ crystal versus temperature
in a large temperature range. 
The solid line results from a fit with the Curie-Weiss law. A field of 1~mT is applied along 
a diagonal of the cubic crystal structure. The inset shows the low temperature range of the data.} 
\label{Er2Ti2O7_these_inv_suscep}
\end{figure}

In Fig.~\ref{Er2Ti2O7_these_inv_suscep} we display our result for the inverse of the static
susceptibility versus temperature in a large temperature range. The Curie-Weiss law
provides a good description of our data above 30~K. The fit gives  for the Curie-Weiss 
temperature
$ \theta_{\rm CW} = -17.5 \, (3)$~K
and $C = 3.73 \, (4)$~K. This means that $m_{\rm para} = 9.55 \, (10) \, \mu_{\rm B}$, 
in agreement with the result for an isolated Er$^{3+}$ ion.
Because  $ \theta_{\rm CW}$ is negative, the dominant exchange interactions are antiferromagnetic.
For comparison, fitting data recorded between 20 and 50~K on a powder sample in a field of 1~mT, Bramwell {\it et al.}\cite{Bramwell00} found values of
$ \theta_{\rm CW} = -22.3 \, (3)$~K
and $m_{\rm para} = 9.34 \, (9) \, \mu_{\rm B}$,
For data recorded between 50 and 300~K in a field of 50~mT these authors obtain
$ \theta_{\rm CW} = -15.93 \, (3)$~K
and $m_{\rm para} = 8.936 \, (4) \, \mu_{\rm B}$,
the latter at variance with the result for an isolated Er$^{3+}$ ion. From our result, we compute for the frustration 
index\cite{Ramirez01} $f\equiv |\theta_{\rm CW}|/T_{\rm N} = 14$. Since $f\gg 1$, Er$_2$Ti$_2$O$_7$ is a
strongly frustrated magnet. In addition, assuming the Er$^{3+}$ magnetic moments
to interact through a simple nearest-neighbor Heisenberg interaction with exchange integral
${\mathcal I}$ (${\mathcal I} > 0$), {\em i.e.}
\begin{eqnarray}
{\mathcal H} = {{\mathcal I} \over 2} \sum_{i,j, i\neq j} {\bf J}_i \cdot {\bf J}_j
 = {\mathcal I}\sum_{\langle i,j \rangle} {\bf J}_i \cdot {\bf J}_j,
\label{Results_bulk_sus_Weiss_3}
\end{eqnarray}
the molecular-field approximation predicts
\begin{eqnarray}
{\mathcal I} = {3 \, k_{\rm B} |\theta_{\rm CW}| \over  z_{\rm nn} J(J+1)}.
\label{Results_bulk_sus_Weiss_4}
\end{eqnarray}
We denote as $z_{\rm nn}$ the number of nearest neighbor Er$^{3+}$ ions to a given Er$^{3+}$ ion. 
In our case $z_{\rm nn}$ = 6. From the measured $\theta_{\rm CW}$ value and taking into account
that $J=15/2$, we compute ${\mathcal I}/k_{\rm B} = 0.138 \, (2)$~K.

\begin{figure}
\includegraphics[width=0.4\textwidth]{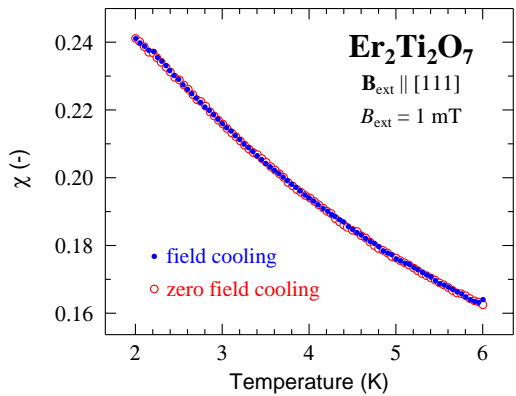}
\caption{(color online) 
The magnetic susceptibility of an Er$_2$Ti$_2$O$_7$ crystal versus temperature in the 
low temperature region. The same magnetic response is observed using either the field 
or zero-field cooling procedure, fc and zfc respectively. A field of 1~mT is applied along a 
diagonal of the cubic crystal structure. In the zfc mode the sample was warmed to 300~K and then 
cooled to 2~K before applying the magnetic field. In the fc mode the field was applied at
300~K and the sample was subsequently cooled down.} 
\label{Er2Ti2O7_these_suscep_Yann}
\end{figure}

We have also measured the susceptibility for $2.0 < T < 6.0$~K under a field of 1~mT applied 
along a $[111]$ axis using two protocols; 
see Fig.~\ref{Er2Ti2O7_these_suscep_Yann}. Contrary to a previous 
report,\cite{Bramwell00} we do not observe any history dependent effect at $T \leq 3.2$~K.
Hence, there is no spin-glass-like irreversible effect for our Er$_2$Ti$_2$O$_7$ crystals.

\subsection{Specific heat in zero magnetic field} 
\label{Results_bulk_Cp_zero_field}

Here we present and discuss zero-field specific heat data recorded for Er$_2$Ti$_2$O$_7$ crystals.
It is well known that they may lead to a characterization of  the low energy 
magnetic modes, detect indirectly a dynamical magnetic component in the ordered state,  determine 
the universality class of the system under study, and gauge a possible residual entropy
at low temperature.

\begin{figure}
\includegraphics[width=0.4\textwidth]{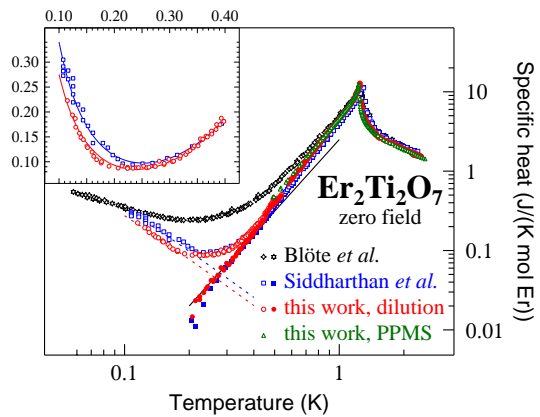}
\caption{(color online) 
Low temperature specific heat of Er$_2$Ti$_2$O$_7$. The open symbols show our experimental data compared to literature results from Bl\"ote {\it et al.}\cite{Blote69} and Siddharthan {\it et al.}\cite{Siddharthan99} Note the large overlap between our results obtained with the PPMS and the 
home-made calorimeter inserted in a dilution refrigerator.
The filled symbols present the electronic specific heat deduced from the data of Siddharthan {\it et al.} and this work. The dashed lines show the contributions from the nuclear specific heat 
and the full line results from a fit of the low temperature electronic specific heat to the  $C_{\rm elec} ={\mathcal B} T^3$ law, with $\mathcal{B}$ = 
2.50\,(15)~J\,K$^{-4}$\,mol$^{-1}$.
The inset displays the very low temperature details. The solid lines result from fits as explained in the main text.} 
\label{Er2Ti2O7_these_chaleur_specifique_low_T}
\end{figure}

In Fig.~\ref{Er2Ti2O7_these_chaleur_specifique_low_T} we display our data, in the
low temperature regime.   
While our results are in reasonable agreement with data published by  
Siddharthan {\it et al.},\cite{Siddharthan99} they differ drastically at low temperature with
the ones of  Bl\"ote {\it et al.}\cite{Blote69} 
We do not understand the origin of the large difference with the data of  Bl\"ote {\it et al.}
In the following we shall focus on the analysis of 
our data and the ones of Siddharthan {\it et al.}
Note that here we have not considered the results of
Sosin {\it et al.}\cite{Sosin10} since they do not extend to very low temperature.

In the low temperature range shown in
Fig.~\ref{Er2Ti2O7_these_chaleur_specifique_low_T} the specific heat arises from three origins: the contribution from the nuclei with non-zero spins at low
temperature, then the low energy magnon modes at higher temperature 
and the critical fluctuations around $T_{\rm N}$.
As a start we shall focus on the first two contributions, {\em i.e.}
the specific heat of nuclear and magnon origins, $C_{\rm n}$ and $C_{\rm sw}$, respectively. 

We begin with providing a theoretical background for these contributions, focusing first on the nuclear one. $C_{\rm n}$ arises from the $^{167}$Er nuclear magnetic moments ($^{167}$Er is the only non-spinless isotope of Er with 23\% natural abundance), since the contribution of the two Ti isotopes is negligible, as in the case of Tb$_2$Ti$_2$O$_7$.\cite{Yaouanc11a} 
Contrary to Tb$_2$Ti$_2$O$_7$ the quadrupole interaction is not negligible compared to the Zeeman interaction. 
This is due to the fact that the quadrupole moment $Q_{167}$ of $^{167}$Er is larger than that of $^{159}$Tb (3.565 vs 1.432~barns) and the gyromagnetic ratio $\gamma_{167}$ of $^{167}$Er is much smaller, in absolute value, than that of $^{159}$Tb ($-7.7157$ vs 64.31~Mrad\,s$^{-1}$\,T$^{-1}$); see Ref.~\onlinecite{Harris01}. The Zeeman and quadrupolar Hamiltonians are written
\begin{eqnarray}
{\mathcal H}_{{\rm Zee}} & = & 
-\hbar \gamma_{167} {\bf I} \cdot {\bf B}_{\rm hyp}
\end{eqnarray}
and
\begin{eqnarray}
{\mathcal H}_{{\rm Q}} & = &\hbar \omega_{\rm Q}
  \left [3 I_z^2 -  I(I+1)
 \right ],
\end{eqnarray}
respectively. In these equations, ${\bf I}$ is the $^{167}$Er spin operator ($I$ = 7/2) and $\hbar \omega_{\rm Q}$ = ${e Q_{167} V_{zz}\over 4 I \left (2 I -1 \right )}$ where $V_{zz}$ is the principal component of the electric field gradient tensor acting on the rare earth nucleus with $z$ being as before the local three-fold axis. The symmetry at the rare earth site 
imposes the electric-field gradient to be axial. Because the Er$^{3+}$ ordered magnetic moments are (nearly) perpendicular to $z$ we shall also take ${\bf B}_{\rm hyp}$ perpendicular to this axis. As usual $\hbar$ and $e$ stand for the Dirac constant and the proton electric charge, respectively.

The nuclear energy levels are determined after diagonalization of the Hamiltonian ${\mathcal H}_{\rm n}$ = ${\mathcal H}_{{\rm Zee}} + {\mathcal H}_{{\rm Q}}$ and $C_{\rm n}$ is readily obtained. ${\mathcal H}_{\rm n}$ depends on two parameters, $B_{\rm hyp}$ and $V_{zz}$. While an estimate for $V_{zz}$ is provided in Appendix~\ref{Vzz}, $B_{\rm hyp}$ will be a fitting parameter.

The other contribution to the low temperature specific heat arises from magnons. Low energy magnons have indeed been observed in neutron scattering experiments.\cite{Ruff08} 
The dispersion relation $\hbar \omega ({\bf q})$ for their lowest energy branch is needed to compute  $C_{\rm sw}$. An approximate expression valid at small wavevectors is
\begin{eqnarray}
\hbar^2 \omega^2 ({\bf q}) = \hbar^2 \omega^2 (q) = \Delta^2_{\rm sw} + 
\hbar^2 v_{\rm sw}^2 q^2.
\label{magnon_lin_2_bis}
\end{eqnarray}
Here $\Delta_{\rm sw}$ is the gap energy of the magnon spectrum at the zone center and $v_{\rm sw}$ is the magnon velocity. 
We note that a dispersion relation has recently been proposed for Er$_2$Ti$_2$O$_7$ in the framework of linear spin-wave theory.\cite{Savary12a} The applicability of this theory in frustrated systems might be questionable as recently discussed in the case of the triangular lattice.\cite{Zheng06a} Still, the model of Ref.~\onlinecite{Savary12a} leads to an anisotropic dispersion relation. 
The resulting specific heat depends on a single magnon velocity which is the geometrical mean of the three magnon velocities along orthogonal axes. In our model it corresponds to $v_{\rm sw}$.

When $\Delta_{\rm sw}$ is negligible, the magnon specific heat $C_{\rm sw}$ can be computed in the temperature range where only small wavevectors are at play, i.e. when Eq.~\ref{magnon_lin_2_bis} applies. 
The expected $T^3$ law for $C_{\rm sw}$ is derived:  
\begin{eqnarray}
C_{\rm sw} =  {\mathcal A} T^3 \ \ {\rm with} \ \    
{\mathcal A} = {\pi^2 \over 120}{\mathcal N}_{\rm A}
{k^4_{\rm B} a^3 \over \hbar^3 v^3_{\rm sw}},
\label{magnon_lin_4}
\end{eqnarray}
where ${\mathcal N}_{\rm A}$ is Avogadro's constant.
This result only holds at sufficiently low temperature.

Having established the theoretical background, we now perform the specific heat data analysis. 
We shall do it in two steps.

We first attempt to determine whether a $T^3$ behavior can be observed. A fit to the measured 
specific heat at the lowest temperatures for which the magnon contribution should be 
negligible enables us to estimate $C_{\rm n}(T)$ and then
to subtract it from the measured heat capacity. The resulting electronic heat capacity, $C_{\rm elec}$, 
is presented in the main panel of Fig.~\ref{Er2Ti2O7_these_chaleur_specifique_low_T} for our data and the ones of 
Siddharthan {\it et al.}
It follows nicely a $T^3$ law in a restricted temperature range, but deviates 
above $\approx T_{\rm N}/2.5$, in contrast to published 
results.\cite{Champion03,Sosin10} This observation justifies to identify 
$C_{\rm elec}$ with $C_{\rm sw}$.
The $T^3$ behavior is not expected to be seen at low temperature if the energy gap 
is appreciable. The effect of the gap might be seen around $T=0.2$~K; see 
Fig.~\ref{Er2Ti2O7_these_chaleur_specifique_low_T}. However, $C_{\rm elec}$
becomes very small at that temperature and difficult to  measure
as reflected by the distribution of the $C_{\rm elec}$ data.
Hence, we cannot determine whether a gap is present from this plot.
Numerically, since we can identify ${\mathcal B}$ given in the caption of Fig.~\ref{Er2Ti2O7_these_chaleur_specifique_low_T} with 
${\mathcal A}$ of Eq.~\ref{magnon_lin_4}, we get
$v_{\rm sw}$ = 86\,(2)~m\,s$^{-1}$.

The second step for the interpretation of the specific heat data consists in fitting the measured specific heat to the sum 
$C_{\rm n} + C_{\rm sw}$. This sum depends on three parameters, $\Delta_{\rm sw}$, $v_{\rm sw}$ and $B_{\rm hyp}$. The fit is shown in the inset of 
Fig.~\ref{Er2Ti2O7_these_chaleur_specifique_low_T}.
Its temperature range is restricted on the high temperature side because of the
requirement that only the low energy magnons
determine the value of the integral. Two solid lines are drawn in
the figure since the two data sets are slightly different at low temperature.
Both sets can be fit to a range of gaps extending from nearly zero to an upper
bound. We find $\Delta_{\rm sw}/k_{\rm B}\leq$ 0.5\,(1)~K for both data sets. 
This is consistent with the value of Sosin {\it et al.}\cite{Sosin10}
We also derive $v_{\rm sw}$ = 84\,(2) and 82\,(2)~m\,s$^{-1}$ and $B_{\rm hyp}$ = 345\,(10) 
and 305\,(5)~T for the Siddharthan {\it et al.} data and our data, respectively. Note that $B_{\rm hyp}$ depends very little on the actual value chosen for $V_{zz}$.

We now discuss these results, starting with the bound on the spin gap energy.
This bound is really small and might be surprising at a first sight given the strong magnetic anisotropy of Er$_2$Ti$_2$O$_7$. However, the Er$^{3+}$ magnetic moment lies at a polar angle of nearly 90$^\circ$ with respect to the local three-fold axis. This angle is imposed by the relatively strong crystal field interaction. There is still a continuous degree of freedom for the azimuthal angle. The magnetic order breaks this rotational symmetry and $\Delta_{\rm sw}$ is a measure of the residual anisotropy energy. We note that the upper bound value for $\Delta_{\rm sw}$ is in the expected range if it arises from the dipole interaction between the Er$^{3+}$ magnetic moments.

We examine now the magnon velocity value and tentatively relate it with the exchange integral introduced in Eq.~\ref{Results_bulk_sus_Weiss_3}. For this purpose we resort to the phenomenological dispersion relation
\begin{eqnarray}
\hbar^2 \omega^2 ({\bf q}) = \hbar^2 \omega^2 (q) = \Delta^2_{\rm sw} + 
\left [  {\mathcal I} z_{\rm nn} J \sin (q d)  \right ]^2,
\label{magnon_lin_2}
\end{eqnarray}
where $d$ is the distance between two magnetic atoms. In fact, had we written $\sin (q^Z d)$ instead of 
$\sin (q d)$, Eq.~\ref{magnon_lin_2} would give the dispersion relation of an 
antiferromagnetic chain running along the $Z$ direction and of lattice parameter $d$ 
for which the number of nearest neighbors is $z_{\rm nn} =2$; see for example Ref.~\onlinecite{Lovesey86a}. 
Here we shall take $d$ as the distance between two Er atoms,
{\em i.e.} $d = a/(2 \sqrt {2})$.
Identifying the small $q$ expansion of Eq.~\ref{magnon_lin_2} with Eq.~\ref{magnon_lin_2_bis} we have
\begin{eqnarray}
 v_{\rm sw} =  {15 \over 2 \sqrt{2}} a  {{\mathcal I} \over \hbar},
\label{magnon_lin_3}
\end{eqnarray}
using $J=15/2$.
This relation leads to ${\mathcal I}/k_{\rm B}$ = 0.118\,(4)~K, a value in reasonable agreement with the one derived from the Curie-Weiss constant; see Sec.~\ref{Results_bulk_sus}.
We shall return to the interpretation of the magnon velocity at the end of 
Sec.~\ref{Results_microscopic_neutron} once a nearest neighbor Hamiltonian consistent with the symmetry of the lattice has been introduced.

It is possible to get interesting information from the $B_{\rm hyp}$ values.
Using the hyperfine constant which is $87 \, (1) \, {\rm T}/\mu_{\rm B}$,\cite{Ryan03} the magnetic moment
at the origin at the field can be obtained. We derive for the moment $4.0\,(2)$ and 
$3.5 \, (1) \, \mu_{\rm B}$ from the Siddharthan {\it et al.}\ measurement and ours, respectively.
The latter value is consistent with the neutron result.\cite{Poole07}
Hence, contrary to Tb$_2$Sn$_2$O$_7$ (Ref.~\onlinecite{Bonville10}) 
there is no reduction of the hyperfine field at the $^{167}$Er nuclei due to electronic spin dynamics. 
This means that the characteristic time 
for the electronic spin-flip is substantially larger than the $^{167}$Er spin-lattice relaxation 
time.\cite{Bertin02}

\begin{figure}
\includegraphics[width=0.4\textwidth]{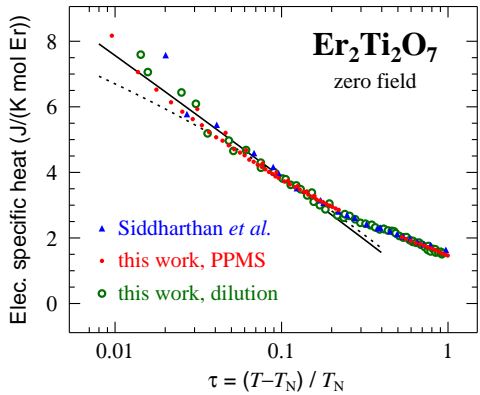}
\caption{(color online) 
Zero-field specific heat of Er$_2$Ti$_2$O$_7$ versus the reduced temperature parameter $\tau$ 
in the paramagnetic regime for three data sets.
The full (dotted) line is the prediction of Eq.~\ref{Results_bulk_Cp_critic} for $\alpha$ = $-0.015$ ($-0.134$). For $\alpha$ = $-0.015$ we find $C_{\rm sh}$
= 1.7\,(1)~J\,K$^{-1}$\,mol$^{-1}$. The critical regime is observed up to $\tau$ $\simeq 0.2$. } 
\label{Er2Ti2O7_sh_exposant_critique}
\end{figure}

We now turn our attention to the critical behavior of the specific heat in the
paramagnetic phase. In Fig.~\ref{Er2Ti2O7_sh_exposant_critique} we display our data using a reduced temperature scale. We expect to observe the usual power law critical behavior:\cite{Fisher64,Kornblit73}  
\begin{eqnarray}
C_{\rm elec}(T) = \frac{C_{\rm sh}}{\alpha} \left[ \left ({T - T_{\rm N} \over T_{\rm N} }\right )^{-\alpha} - 1 \right],
\label{Results_bulk_Cp_critic}
\end{eqnarray}
where $C_{\rm sh}$ is a constant and $\alpha$ the specific heat critical exponent.
By definition, $C_{\rm elec}(T)$ has a  maximum at $T_{\rm N}$. This enables us to
determine  $T_{\rm N}$, as already mentioned in Sec.~\ref{Experimental}. The exponent $\alpha$ is expected to be $\alpha = -0.015$ and $-0.134$ for the three-dimensional XY and Heisenberg magnets, respectively. As seen in Fig.~\ref{Er2Ti2O7_sh_exposant_critique}, Eq.~\ref{Results_bulk_Cp_critic} provides a better account of the data for the XY case, as expected.

\begin{figure}
\includegraphics[width=0.4\textwidth]{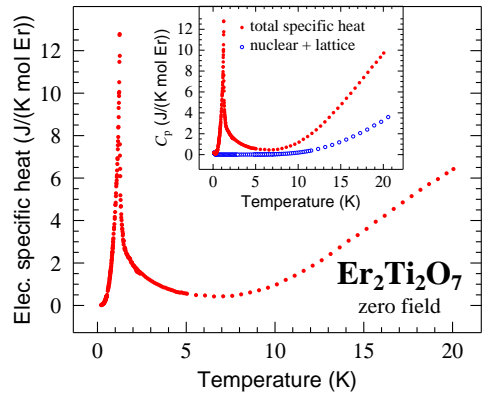}
\caption{(color online) 
The electronic specific heat of Er$_2$Ti$_2$O$_7$ in an extended temperature range.
In the inset are displayed the total specific heat ($C_{\rm p}$) of 
Er$_2$Ti$_2$O$_7$ and the
sum of the estimated nuclear and lattice contributions.} 
\label{Er2Ti2O7_these_chaleur_specifique_global}
\end{figure}

\begin{figure}
\includegraphics[width=0.4\textwidth]{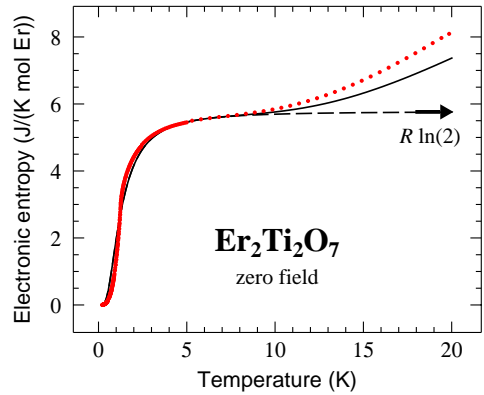}
\caption{(color online) 
The variation of the electronic entropy $\Delta S_{\rm elec}(T)$  of Er$_2$Ti$_2$O$_7$   versus 
temperature $T$.
The solid line is computed assuming two singlet states at energies corresponding 
to 0 and 2.85~K and two doublets at 74 and 85.8~K. A signature of the two doublets 
has been seen by inelastic neutron scattering.\cite{Champion03} The ground  state doublet 
has been split to account for the magnetic order of Er$_2$Ti$_2$O$_7$ 
below $T_{\rm N}$, with the splitting taken as a fitting parameter. Admittedly, this is  
a very rough description of the physics. The dashed line is the result of the computation of $\Delta S_{\rm elec}(T)$
when only the first two levels are taken into account. 
} 
\label{Er2Ti2O7_these_entropie}
\end{figure}

Before leaving this section, we discuss the entropy variation of our system using our 
specific heat results. We have extended the specific heat measurement $C_{\rm p}$ of  
Er$_2$Ti$_2$O$_7$ up to approximately 20~K and measured the specific heat of the isostructural 
non-magnetic compound Y$_2$Ti$_2$O$_7$ in the same temperature range. 
The sum of the contributions from the nuclear moments and
the lattice, the latter being estimated from scaling the Y$_2$Ti$_2$O$_7$ 
result,\cite{Chapuis10} is presented in the inset of 
Fig.~\ref{Er2Ti2O7_these_chaleur_specifique_global}, as well as $C_{\rm p}(T)$.
The resulting $C_{\rm elec}(T)$ obtained from 
subtracting the contributions of the nuclear moments and the lattice to $C_{\rm p}(T)$ is 
displayed in the main frame of Fig.~\ref{Er2Ti2O7_these_chaleur_specifique_global}. 
In Fig.~\ref{Er2Ti2O7_these_entropie} we present $\Delta S_{\rm elec}(T)$, which is the temperature
variation of the electronic entropy obtained by integrating $C_{\rm elec}(T^\prime)/T^\prime$ from 
$0.115$~K, the lowest measured temperature, to the temperature of interest $T$. 
Remarkably, $\Delta S_{\rm elec}(T)$ reaches the $R \ln (2)$ value at approximately 8~K. This is the entropy variation expected for an isolated doublet state. Therefore the residual entropy left as $T \rightarrow 0$ is extremely small if any. For temperatures above 8~K, $\Delta S_{\rm elec}(T)$ keeps increasing owing to the contribution of the excited CEF levels. This is clearly shown in Fig.~\ref{Er2Ti2O7_these_entropie}, where a comparison is made between the results of computations of $\Delta S_{\rm elec}(T)$ either taking into account the first excited CEF levels or neglecting them.

\subsection{Specific heat under an external magnetic field} 
\label{Results_bulk_Cp_field_dependence}

\begin{figure}
\includegraphics[width=0.4\textwidth]{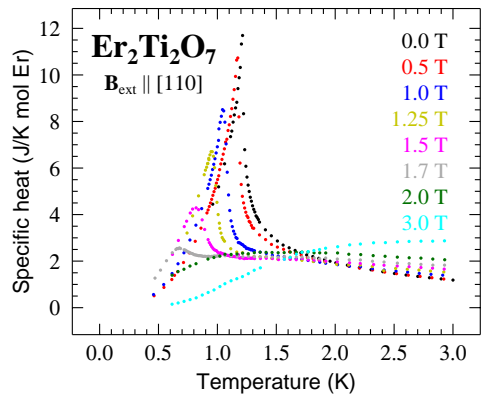}
\caption{(color online) 
Temperature dependence of the specific heat of a Er$_2$Ti$_2$O$_7$ single crystal for
different magnetic field intensities applied along $[110]$.
The maximum of the specific heat peak moves to lower temperatures as the field increases
up to 1.7~T. No peak is observed when the field strength is above 1.7~T.} 
\label{Er2Ti2O7_these_chaleur_specifique_field}
\end{figure}

\begin{figure}
\includegraphics[width=0.4\textwidth]{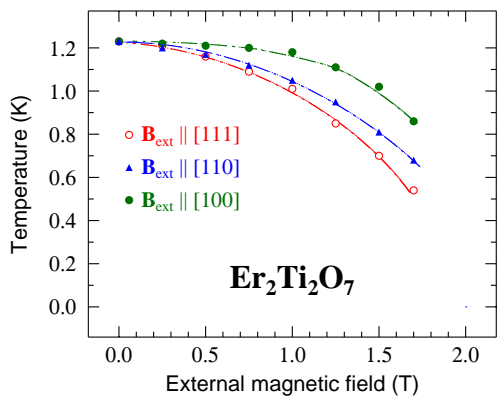}
\caption{(color online) 
The phase diagram derived from specific heat measurements for the three main crystal 
directions of cubic Er$_2$Ti$_2$O$_7$. The dashed-dotted lines are guides to the eye.} 
\label{Er2Ti2O7_these_phase_diagram}
\end{figure}

We have constructed the phase diagram of Er$_2$Ti$_2$O$_7$ in the field-temperature plane using 
specific heat data. Examples of measurements are shown in 
Fig.~\ref{Er2Ti2O7_these_chaleur_specifique_field}.   
For a given external field we have determined the temperature at which the specific heat 
displays a maximum. The position of the maximum as a function of the field intensity for a 
given field orientation relative to the crystal axes is displayed in 
Fig.~\ref{Er2Ti2O7_these_phase_diagram}. 
Our results are consistent with the ones already published,\cite{Ruff08,Sosin10} but only
qualitatively. Note that here we establish the phase diagram for the three main directions 
of a cubic compound.
These data suggest a quantum critical point to be present slightly above 2~T and to be dependent on the field orientation.

\section{Microscopic technique measurements} 
\label{Results_microscopic}

Single crystals of Er$_2$Ti$_2$O$_7$ have been studied by two microscopic 
experimental techniques: positive muon spin relaxation ($\mu$SR) and neutron scattering. We shall first discuss the $\mu$SR results.

\subsection{$\mu$SR} 
\label{Results_microscopic_muon}

\begin{figure}
\includegraphics[width=0.4\textwidth]{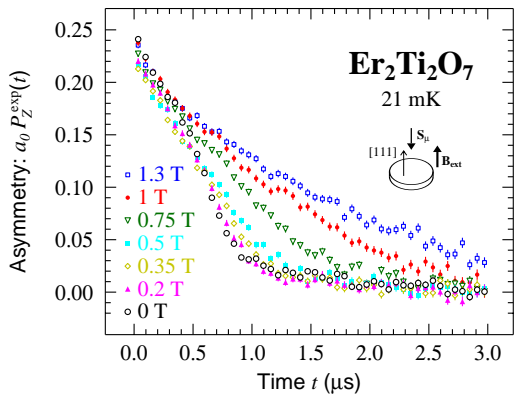}
\caption{(color online) 
$\mu$SR spectra of a Er$_2$Ti$_2$O$_7$ crystal taken at S$\mu$S with the initial muon beam 
polarization parallel to the $[111]$ crystal direction. The spectra were recorded in zero 
and various longitudinal fields as indicated in the figure. 
} 
\label{Er2Ti2O7_these_muon_spectrum}
\end{figure}

In the longitudinal geometry that we have used, a $\mu$SR spectrum recorded in the magnetically
ordered state of a crystal is expected to display either (i) at least one damped oscillation if 
the initial muon beam polarization is not parallel to the spontaneous field at the muon site, or (ii)  
a missing fraction if the oscillation cannot be resolved.\cite{Yaouanc11} 
None of these two 
possibilities was observed at ISIS or S$\mu$S. The zero-field spectrum recorded at 21~mK is displayed in Fig.~\ref{Er2Ti2O7_these_muon_spectrum}; the spectral shape shows little change up to $\approx$ 0.5~K.
The same type of spectra was also observed for 
different orientations of the initial muon beam polarization relative to the crystal axes. 
Hence, the absence of oscillation cannot be attributed to the initial muon beam polarization
which would be parallel to the  internal field. This situation would moreover be unexpected because of the four equivalent symmetry $\langle 111 \rangle$ axes at the Er$^{3+}$ site. In addition to the absence of oscillation,
the shape of the zero-field spectra is extremely unusual. Indeed, the spectral slope is quasi-constant up to $\approx$~0.4~$\mu$s, then it increases and
eventually monotonically decreases above $\approx$~0.7~$\mu$s. This kind of
behavior drastically differs from the usual relaxation spectra, which are 
characterized by a monotonically decreasing slope, as e.g. in the common exponential relaxation. It does not either remind the shape of spectra associated with a static or quasistatic field distribution at the muon. At this stage we cannot actually conclude whether the spectral shape is mainly influenced by a static field distribution or dynamical effects. We shall come back to this point when commenting the data recorded in applied fields.

Since no spontaneous muon precession is observed, it is tempting to make an analogy with Tb$_2$Sn$_2$O$_7$
for which there is also no detected oscillation in the magnetically ordered state.\cite{Dalmas06}
This analogy would be completely justified if the zero-field spectrum was exponential. However, we have just remarked that it is obviously not the case. It is not
Gaussian either, as first suggested.\cite{Lago05} A close look at the data in Ref.\ \onlinecite{Lago05}
shows that the zero-field relaxation is in fact consistent with the one we observe.

Before discussing further the significance of our result, it is worthwhile to  consider the longitudinal field spectra shown in
Fig.~\ref{Er2Ti2O7_these_muon_spectrum}. First note that with the considered field intensities the compound
has {\it a priori} not crossed the phase diagram boundary displayed in  
Fig.~\ref{Er2Ti2O7_these_phase_diagram}. While the application of fields below 0.5~T has little influence on the spectral shape, it is not the case for higher fields. Interestingly, for the highest fields shown in the figure, the spectra tend to be described by an exponential function.\footnote{For instance the spectrum measured at 21~mK under a field of 1.3~T can nicely be fitted to a stretched exponential relaxation function $a_0\exp[-(\lambda_Z t)^\beta]$ with $\beta$ = 0.86\,(3).} This implies that the muon repolarization is not at the origin of the field dependence of the spectra and that dynamics is mainly influencing the muon response in Er$_2$Ti$_2$O$_7$. The time scale of this dynamics can be roughly estimated. For this purpose we have recourse to the Lorentzian field dependence of the exponential relaxation rate $\lambda_Z$ in the motional narrowing, {\sl i.e.} fast fluctuation limit.\cite{Redfield57,Yaouanc11} In this model the relaxation rate in a field $B_{\rm ext}$ is such that $\lambda_Z(B_{\rm ext})/\lambda_Z(B_{\rm ext} = 0)$ = 1/2 for $B_{\rm ext}$ = $B_{1/2}$ = $1/(\gamma_\mu \tau_{\rm f})$. Here $\tau_{\rm f}$ is the fluctuation time of the spins and $\gamma_\mu =851.615 \, {\rm Mrad} \, {\rm s}^{-1} \, {\rm T}^{-1}$ is the muon 
gyromagnetic ratio. Taking 1~T as an order of magnitude for $B_{1/2}$ we find
$\tau_{\rm f}\approx 10^{-9}$~s. 

We have therefore established that the muons are probing dynamical fields. No oscillation is detected in zero-field because the mean field at the muon site does not keep a constant value for a time sufficiently long for a muon spin precession
to be observed.\cite{Dalmas06} We denote this field as ${\bf B}_{\rm fluc}$. It arises from the dipole interaction of the muon magnetic moment with the Er$^{3+}$ magnetic moments. Given the size of the Er$^{3+}$ magnetic moment, we estimate $B_{\rm fluc}$ in the range 0.1-0.2~T. 

With our knowledge for $\tau_{\rm f}$ and $B_{\rm fluc}$, we find 
$\tau_{\rm f} \gamma_\mu B_{\rm fluc}=  B_{\rm fluc}/B_{\rm ext} \ll 1$. For the simple model of 
${\bf B}_{\rm fluc}$ flipping from parallel to antiparallel to an axis perpendicular to initial muon polarization ${\bf S}_\mu$, we 
would expect the zero-field relaxation to be exponential,\cite{Yaouanc11} as found for  
Tb$_2$Sn$_2$O$_7$. However, experimentally this is not the case. This is not
surprising given the fact that the magnetic diffraction profiles are complex. Their shape is
ascribed to the coexistence of long and short-range dynamical correlations; see our discussion
of  Fig.~\ref{Er2Ti2O7_PRL_101_147205_fig2f_reworked}. To the distribution of correlation lengths
must correspond distributions of spin-spin and spin-lattice relaxation rates. This may explain
the exotic shape of the zero-field $\mu$SR relaxation function.

Relative to Tb$_2$Sn$_2$O$_7$, the fluctuations probed by $\mu$SR in  Er$_2$Ti$_2$O$_7$ are slower 
by roughly an order of magnitude. From neutron spin-echo measurements it is known that in 
the ordered state of Tb$_2$Sn$_2$O$_7$ spin correlations near the zone center are static\cite{Chapuis07} within the technique time scale, 
while they are dynamical in nature far outside the center of the zone.\cite{Rule09b} It would be worthwhile to 
examine the dynamics of the magnetic correlations in  Er$_2$Ti$_2$O$_7$ with the neutron spin-echo technique. 

From the analysis of the nuclear specific heat in Sec.~\ref{Results_bulk_Cp_zero_field}, it was 
inferred that the ratio of $\tau_{\rm f}$ to the nuclear spin-lattice relaxation time was larger in Er$_2$Ti$_2$O$_7$ than in Tb$_2$Sn$_2$O$_7$. Assuming the spin-lattice relaxation times in the two compounds to be similar, the nuclear specific heat data are consistent with the results of the analysis of the $\mu$SR spectra.

\subsection{Neutron scattering} 
\label{Results_microscopic_neutron}

\begin{figure}
\center
\includegraphics[angle=270,totalheight=0.13\textheight]{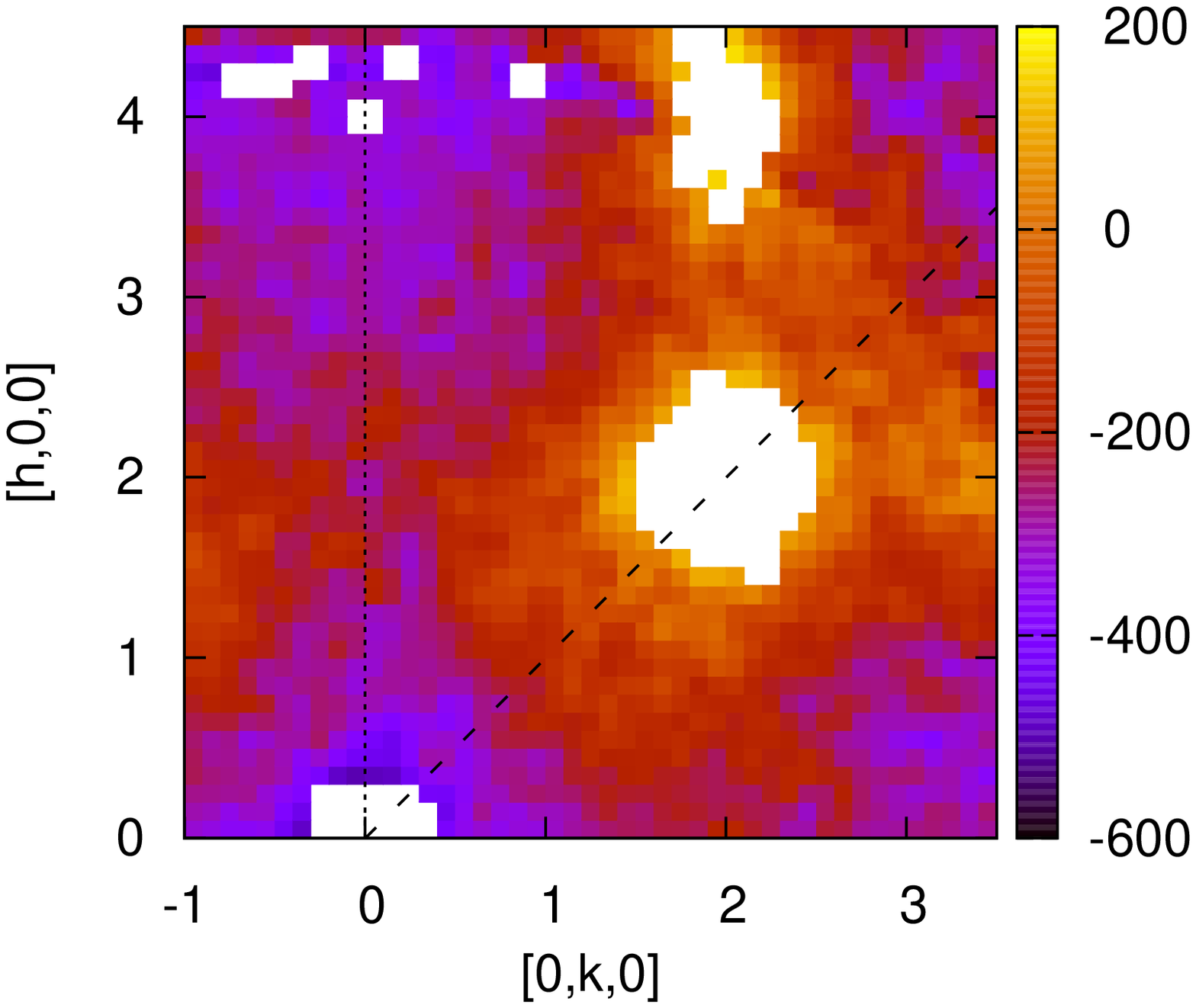}
\includegraphics[angle=270,totalheight=0.13\textheight]{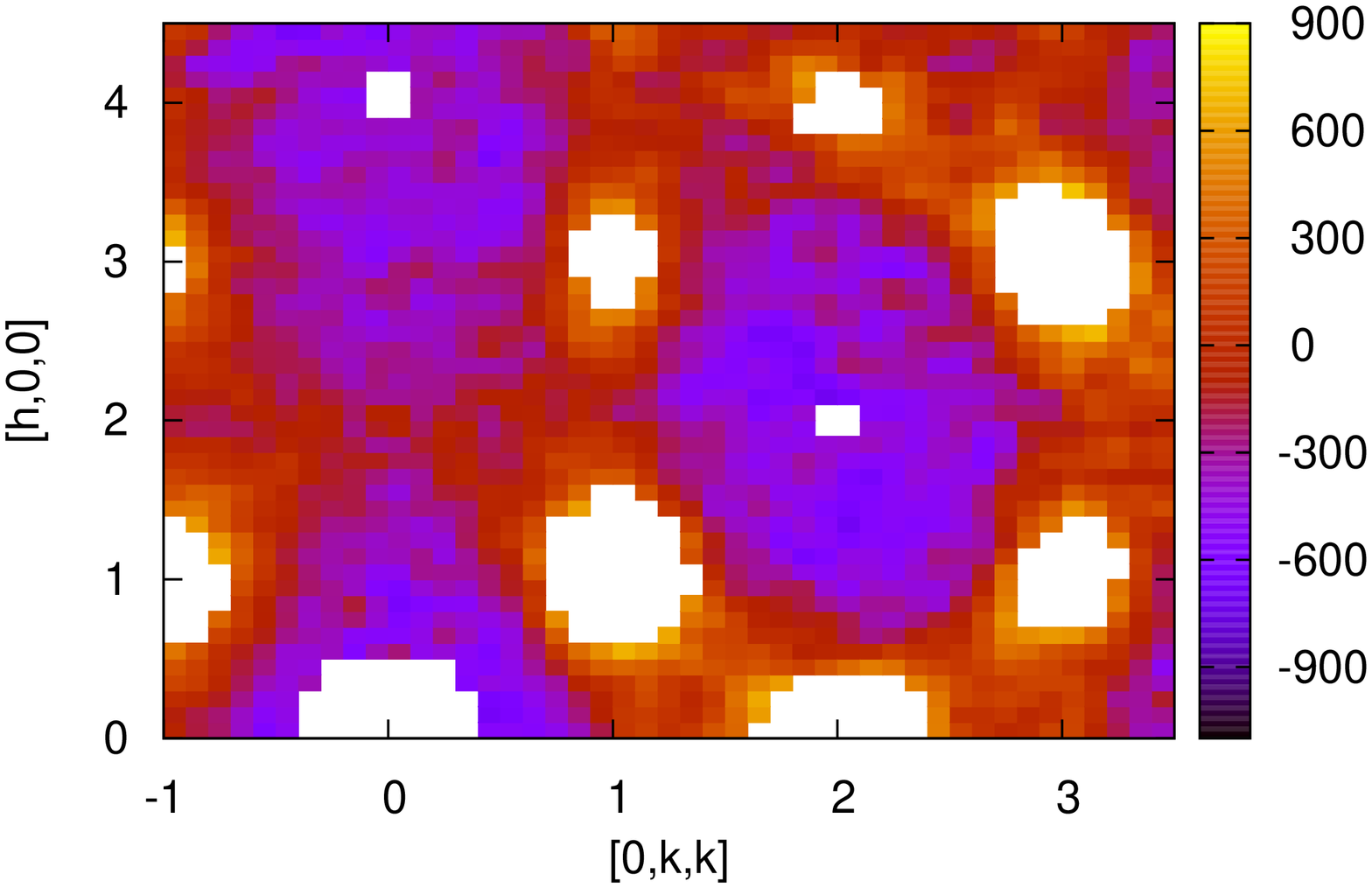}\\
\includegraphics[angle=270,totalheight=0.13\textheight]{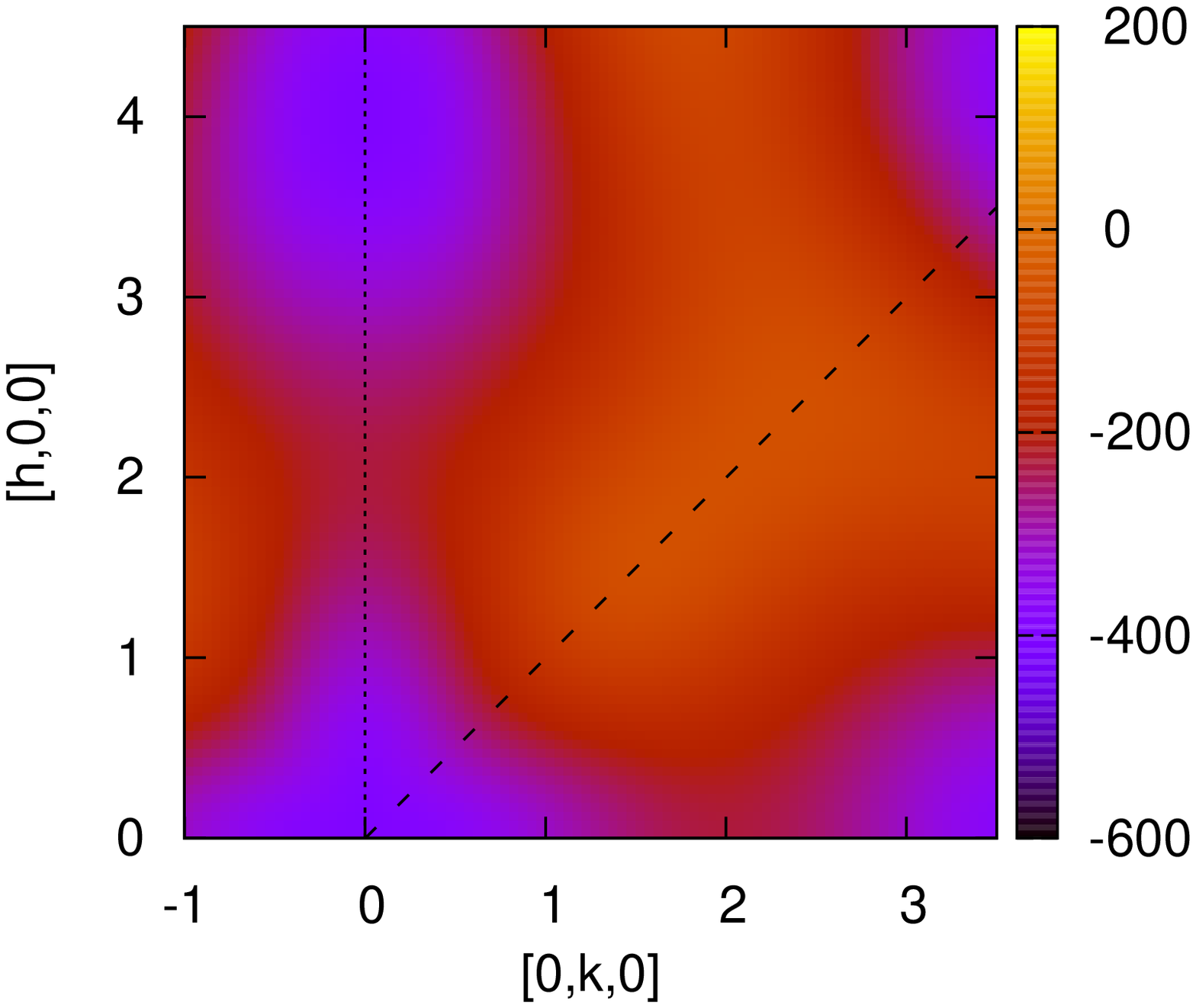}
\includegraphics[angle=270,totalheight=0.13\textheight]{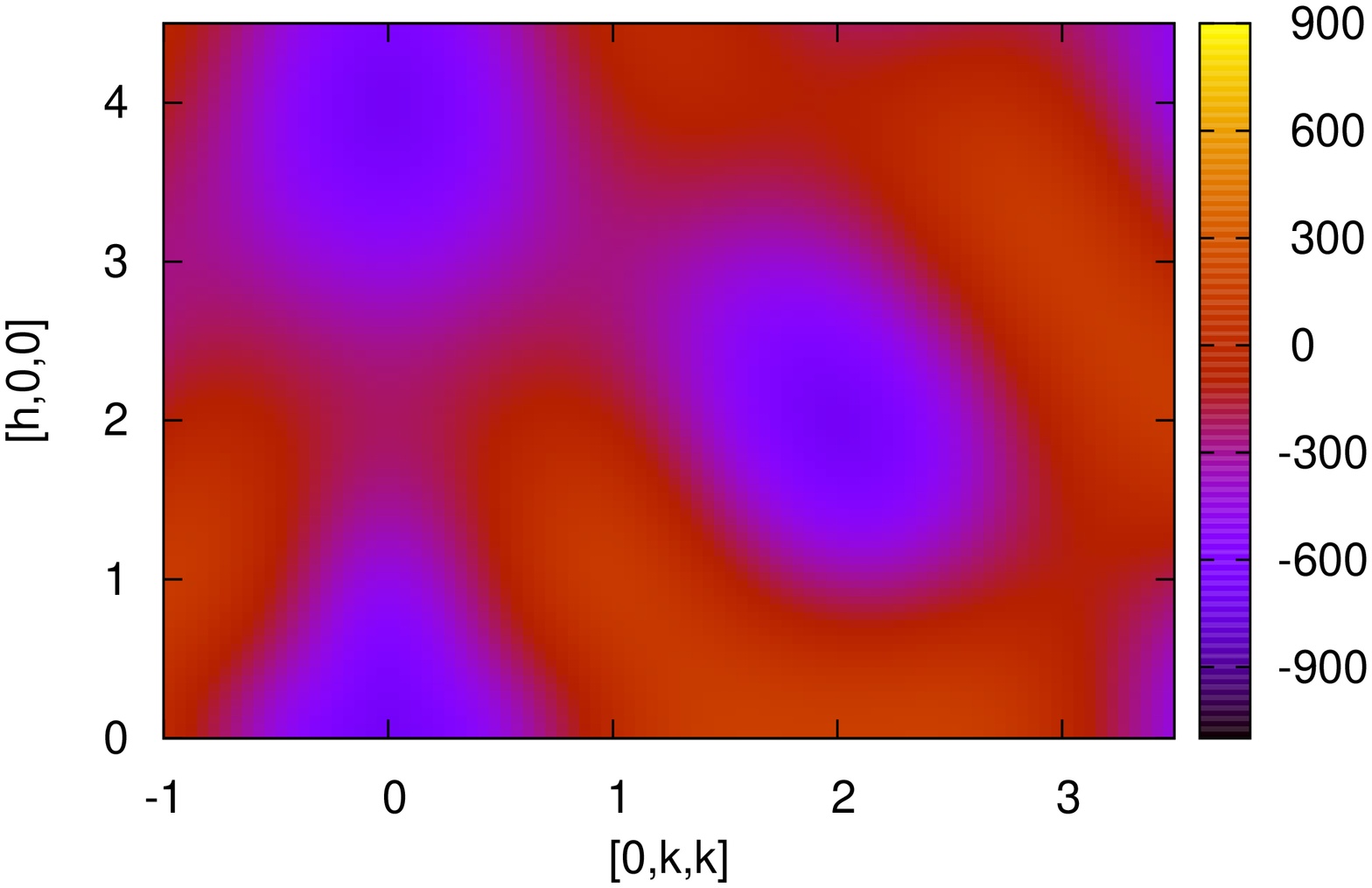}
\caption{(color online) 
Top two panels: magnetic diffuse neutron scattering intensity recorded for a crystal of Er$_2$Ti$_2$O$_7$ in the reciprocal $(h,k,0)$ and $(h,k,k)$ planes at 2.00\,(3) and 1.47\,(3)~K respectively. The positions in the reciprocal lattice are in $2 \pi/a$ units, where $a$ is the lattice parameter of the cubic unit cell. These maps are obtained as explained in the main text.
To preserve the maps appearance, pixels with off scale intensities, e.g pixels influenced by Bragg reflections and critical scattering, as well as pixels located near the origin of the reciprocal lattice have been graphically eliminated: they are represented in white color. 
Bottom two panels: $(h,k,0)$  and $(h,k,k)$ magnetic correlation maps computed with the tetrahedron model explained in the main text. 
The comparison between the theoretical and experimental maps 
displayed above enables us to derive information on the Er$_2$Ti$_2$O$_7$ interaction constants. The lines drawn in the $(h,k,0)$ maps indicate the position of the cuts shown in Fig.~\ref{Er2Ti2O7_these_maps_cut}.} 
\label{Er2Ti2O7_these_maps}
\end{figure}

Paramagnetic correlations were studied by diffuse neutron scattering in Er$_2$Ti$_2$O$_7$ crystals. Two scattering planes were investigated: $(h,k,0)$ and $(h,k,k)$ at 2.00\,(3) and 1.47\,(3)~K, respectively. For the first (second) one a graphite (copper) monochromator was used delivering neutrons of wavelength 2.377 (1.275)~\AA. 
No energy analysis of the scattered beam was performed. In order to deal with magnetic correlations only, additional maps were recorded at 50~K for both geometries and they were subtracted from the corresponding low temperature counterparts. The resulting two maps were divided out by the square modulus of the Er$^{3+}$ magnetic form factor.
\footnote{The form factor of the Er$^{3+}$ ion was taken from  
http://www.neutron.ethz.ch/research/resources/formfactor.}
The effect of the form factor is modest: no more than 20\% on the vast majority of the data.
The resulting experimental data are displayed in Fig.~\ref{Er2Ti2O7_these_maps}.  It is to be noted that the intensities are negative for large regions in the two planes. This reflects the fact that the wavevector independent scattering associated with the CEF states of the Tb$^{3+}$ ions is larger at 50~K than at low temperature.

Before attempting a quantitative analysis of the maps, a qualitative discussion
is worthwhile. We first note the almost vanishing $(2,2,2)$ spot. This means that the ferromagnetic correlations are negligible.
We observe a hexagonal scattering loop in the plane $(h,k,k)$ around the $(2,2,2)$ position. Such a type of scattering is reminiscent of the intensity measured in the cubic spinel ZnCr$_2$O$_4$ (Ref.~\onlinecite{Lee02}), in which the Cr ions also form a lattice of corner sharing tetrahedra. However, in the latter case
the loop is in the plane $(h,k,0)$ and centered around $(2,2,0)$. The scattering properties are therefore quite different for the 
two compounds. This reflects the difference in 
magnetic symmetry. The origin of the loops observed in ZnCr$_2$O$_4$, which were originally interpreted in terms
of weakly interacting hexagonal spin clusters, is now taken as the signature of extended exchange
interactions for spin-ice and isotropic systems.\cite{Yavors08,Conlon10} In the following we show 
that the scattering loop in Er$_2$Ti$_2$O$_7$ can basically be taken as a fingerprint of the 
properties of the exchange interactions within a single tetrahedron.

The discussion of the experimental results will be carried out in two steps.
We shall first evaluate the magnetic correlation length at the temperature
of the measurements and then analyze the maps using a four-spin Hamiltonian.

\subsubsection{Magnetic correlation length} 
\label{Results_microscopic_neutron_length}

\begin{figure}
\includegraphics[width=0.4\textwidth]{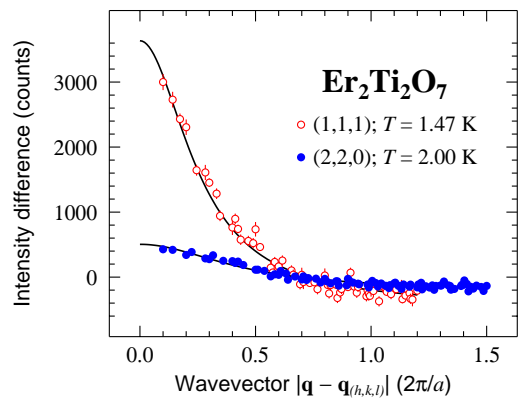}
\caption{(color online).
Magnetic scattering intensity versus wavevector in the vicinity of the two reciprocal lattice
positions ${\bf q}_{(h,k,l)}$ where $(h,k,l)$ = (1,1,1) and (2,2,0), respectively.
The wavevector unit is $2 \pi /a$ where $a$ is the cube edge.
The data are obtained from the intensity differences shown in Fig.~\ref{Er2Ti2O7_these_maps}
by averaging the data at reciprocal space points located at equal
distance from the two lattice positions. Two requirements were considered
when choosing these two positions: they do not lie
close to the boundary of the maps and the magnetic intensity is 
relatively important. The lines are results from fits of
Eq.~\ref{Results_microscopic_neutron_1} to the data.}
\label{Er2Ti2O7_correlation_length}
\end{figure}

Here we
determine the correlation length of the critical magnetic correlations. For this purpose
we consider the scattered intensity measured in the vicinity of the reciprocal positions ${\bf q}_{(h,k,l)}$ = 
${\bf q}_{(2,2,0)}$ and ${\bf q}_{(1,1,1)}$ at $T=2.00$ and $1.47$~K, respectively; 
see Fig.~\ref{Er2Ti2O7_correlation_length}. 
This critical scattering intensity is 
described by the sum of a Lorentzian function and a constant:
\begin{eqnarray}
{\mathcal L} (|{\bf q} - {\bf q}_{(h,k,l)}|) 
= {I_{\rm L} \over 1 + |{\bf q} - {\bf q}_{(h,k,l)}|^2/\kappa_{\rm m}^2} + I_0,
\label{Results_microscopic_neutron_1}
\end{eqnarray}
where $\kappa_{\rm m}$ is the inverse of the magnetic correlation length. The parameter
$I_{\rm L}$
accounts for the magnitude of the Lorentzian, while $I_0$ refers to a neutron
intensity which is not related to critical scattering.  Since at the
temperature of experiments, the width of the critical magnetic scattering curve
is much larger than the instrumental resolution, the convolution of
Eq.~\ref{Results_microscopic_neutron_1} by the resolution function is
unnecessary. The fits shown in Fig.~\ref{Er2Ti2O7_correlation_length} yield
the magnetic correlation lengths $\xi_{\rm m}$ = $\kappa_{\rm m}^{-1}$ = 
3.6\,(2) and 6.6\,(5)~\AA\ for the (2,2,0) and (1,1,1) reflections measured at 2.00 and 1.47~K, respectively. As expected,  $\xi_{\rm m}$ shoots up as the sample is cooled toward the transition.
These two values are comparable with the Er$^{3+}$-Er$^{3+}$ 
ion distance $d =3.56$~\AA. 
Hence the analysis of the experimental maps shown in Fig.~\ref{Er2Ti2O7_these_maps} can be performed considering the spin correlations within a single tetrahedron. 
This is the basis for our quantitative interpretation which is exposed below. 

\subsubsection{Analysis of the diffuse scattering maps}
\label{Results_microscopic_neutron_Analysis}

While our analysis of the magnetic scattering intensity in the vicinity of reciprocal lattice positions
at low temperature shows that the measured wavevector dependence probes short-range correlations,  
a wavevector independent scattering is also observed; see Fig.~\ref{Er2Ti2O7_correlation_length}. 
This scattering reflects local physics, for example of crystal-field nature.
Denoting $M({\bf q})$ a measured magnetic scattering map, we write
\begin{equation}
M({\bf q}) = S_{\rm shift} + S_{\rm scale}\, N({\bf q}),
\label{Results_microscopic_neutron_Analysis_1_1}
\end{equation}
where $S_{\rm shift}$ accounts for the wavevector independent scattering, $S_{\rm scale}$ gives the scale
of the wavevector dependent magnetic intensity and $N({\bf q})$ is the prediction of our model that we describe now. 

In Sec.~\ref{Survey} we have mentioned that the  Er$^{3+}$ crystal-field ground state
doublet is well isolated since the first excited doublet is at about $74$~K above 
the ground state in temperature units. Therefore for low temperature measurements, such as discussed 
here, it is a reasonable approximation to describe an  Er$^{3+}$ ion as an effective $S=1/2$ spin.
In Appendix~\ref{Tetrahedron} we discuss the physics of a tetrahedron of effective one-half spins  
embedded in a pyrochlore lattice. Assuming that only bilinear spin interactions are relevant, the Hamiltonian 
${\mathcal H}_{\rm t}$ describing the interaction between the four effective spins can be 
written as a linear combination of four invariants; see Eq.~\ref{singletet}.
Their weight is gauged by the four parameters $P_i$, $i\in \{1, 2, 3, 4 \}$.

The wavevector dependent scattering intensity resulting from a single tetrahedron is proportional to\cite{Lovesey86a}
\begin{equation}
N({\bf q}) = \sum_{m}e^{-E_m/(k_{\rm B} T)}f_m({\bf q}),
\label{Results_microscopic_neutron_Analysis_1}
\end{equation}
where
\begin{eqnarray}
& & f_m({\bf q})  =  
\label{Results_microscopic_neutron_Analysis_2}\\[1mm]
& & \sum_{\alpha,\beta}\sum_{i,j}
\sum_n
\left(\delta^{\alpha \beta}-\frac{{q}^\alpha{q}^\beta}{q^2}\right)
\langle m|S_i^\alpha|n\rangle\langle n|S_j^\beta|m\rangle e^{i{\bf q}\cdot ({\bf r}_j
-{\bf r}_i)}.
\nonumber
\end{eqnarray}
The Debye-Waller factor is negligible for the region of the reciprocal 
space investigated and the temperatures at which the measurements were made. 
The symbols $\alpha,\beta$ refer to global Cartesian coordinates in the cubic unit cell. 
The relations between the local and global 
axes are explicitly given in Appendix~\ref{Tetrahedron_geometry}.
The indices $i$ and $j$ stand for the Er ions at the corners of the tetrahedron,
the positions of which are specified by ${\bf r}_i$ and ${\bf r}_j$, and
$|n\rangle$ and $|m\rangle$ refer to two of the sixteen single tetrahedron states, whose energy 
differences lie within the energy range across which the neutron scattering 
is integrated.  The formula in
Eq.~\ref{Results_microscopic_neutron_Analysis_2} does not contain the Er$^{3+}$ form
factor since it has already been divided out for the two presented maps. 

\begin{figure}
\includegraphics[width=0.4\textwidth]{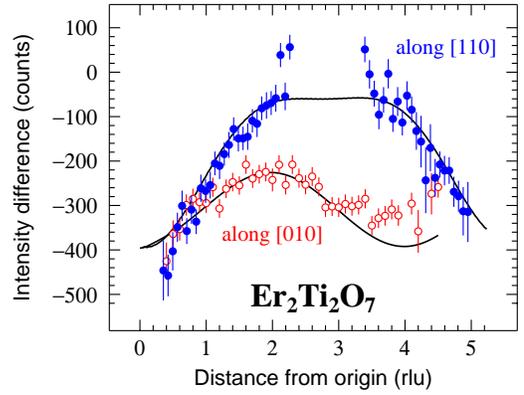}
\caption{(color online) 
Cuts of the $(h,k,0)$ maps shown in Fig.~\ref{Er2Ti2O7_these_maps} along the [1,1,0] and [0,1,0] directions. The circles represent the experimental data and the lines the results of the model. The origin of the horizontal axis is taken at the origin of the reciprocal lattice.} 
\label{Er2Ti2O7_these_maps_cut}
\end{figure}

We have performed a global fit of the two recorded maps. It depends
on two $ S_{\rm shift}$ parameters, one per map, a unique scale parameter 
$S_{\rm scale}$ and the four Hamiltonian parameters $P_i$. 
The best fit shown in Fig.~\ref{Er2Ti2O7_these_maps} is
achieved with the following interaction constants in kelvin units:
\begin{eqnarray}
P_1/k_{\rm B} & = 10.4 \, (6), \ \ \ P_2/k_{\rm B} & = -1.2 \, (5), \cr
P_3/k_{\rm B} & = 3.0 \, (3),  \ \ \ \ P_4/k_{\rm B} & = 8.4 \, (1.3).
\label{Results_microscopic_neutron_Analysis_3}
\end{eqnarray}
The quality of the fit can be assessed from cuts of the experimental and model maps as seen in Fig.~\ref{Er2Ti2O7_these_maps_cut}.

The fitting parameters described above were determined from least square minimization, {\em i.e.} minimizing $\chi^2$ = 
$[1/(N_d-7)]\sum_j (M_j-I_j)^2/\sigma_j^2$ with respect to the fitting parameters. Here $\sigma_j$ is the statistical
uncertainty on the neutron intensity $I_j$ at the $j$th data point in the set of 
$N_d$ data points and $M_j$ 
is the model prediction (a function of the fitting parameters). 
An uncertainty $\delta P_i$ on $P_i$ is obtained by varying 
the neutron intensity by $\partial I_j$ at each of a
large sample of the data points (one in eight) and minimizing $\chi^2$
in order to find 
the variation of each
fitted parameter $\partial P_i$.  Then $\delta P_i$ is estimated using the relation
$\delta P_i = (1/N_d)\sqrt{\sum_j (\sigma_j  \partial P_i/\partial I_j)^2}$.
 
We have so far described the estimate of the statistical uncertainties. Three other origins for systematical uncertainties must be considered.
First our model describes diffuse scattering and therefore pixels which are sizably influenced by critical scattering should be eliminated in the fitting procedure. For this purpose we have tested the influence of different radius cutoffs around the Bragg points, to the $P_i$ values. This cutoff effect introduces the largest parameter uncertainties. 
Then the temperature at which a map is recorded is known with a finite uncertainty. Finally the uncertainty arising from the error bars on the matrix elements $j_{\rm CEF}$ and $t_{\rm CEF}$ (see Eq.~\ref{local_5}) has also been assessed.
The error bars given in Eq.~\ref{Results_microscopic_neutron_Analysis_3} account for these statistical and systematical uncertainties.

It is tempting to extract information from the Curie-Weiss constant using the 
high-temperature susceptibility formula derived Ross {\it et al.}\cite{Ross11} However,
this should not be done since that formula depends on spectroscopic factors
which provide a description of the physics only at low temperature. A high-temperature
expansion for the susceptibility is required to further analyze the data of 
Fig.~\ref{Er2Ti2O7_these_inv_suscep}.   

At this juncture it is of interest to mention a recent work by Savary {\em et al.\@} on Er$_2$Ti$_2$O$_7$.\cite{Savary12a} These authors analyze the spin wave dispersions measured for different orientations of the wave vector. The experiments were performed for temperature (30~mK) and field (3~T) values where Er$_2$Ti$_2$O$_7$ is a polarized paramagnet.\cite{Ruff08} 
The analysis of the data involves a nearest neighbor Hamiltonian related to ours, with a different definition of the Hamiltonian parameters. Using the relations between the two sets of parameters given in Eq.~\ref{Tetrahedron_parameters_6}, and assuming $P_i$ = $2{\tilde P}_i$ we find for their parameters in kelvin units,
\begin{eqnarray}
P_1/k_{\rm B} & = 1.74 \, (1.26), \ \ \ P_2/k_{\rm B} & = -0.44 \, (74), \cr
P_3/k_{\rm B} & = 2.92 \, (34), \ \ \ \ \ P_4/k_{\rm B} & = 9.0 \, (1.0).
\label{Results_microscopic_neutron_Analysis_4}
\end{eqnarray}
These $P_2$, $P_3$ and $P_4$ values are in very good agreement with ours. However an important difference is observed in $P_1$, i.e. the parameter which controls the interaction along the local hard magnetic axis. 
We have simulated the diffuse scattering intensity obtained from this latter set of parameters and compared it with our experimental data. We obtain an acceptable fit of our data since the confidence parameter is $\chi^2$ = 1.44 to be compared with 1.20 with the parameter set in Eq.~\ref{Results_microscopic_neutron_Analysis_3}. 
At this point it must be noted that our data were recorded in zero external field, contrary to those of Ref.~\onlinecite{Savary12a} and that the anisotropy ratio $g_\perp/g_\parallel$ = 4.3 that we have adopted is quite larger than the one chosen by Savary {\em et al.}: $g_\perp/g_\parallel$ = 2.4. 

As already mentioned the dispersion relation of the magnon modes has been computed in the framework of linear spin-wave theory.\cite{Savary12a} Neglecting the gap, an assumption which is justified (see Sec.~\ref{Results_bulk_Cp_zero_field}), we can compute the geometric mean $\bar{v}_{\rm sw}$ of the magnon velocities along the three directions of the Cartesian frame. The expression is:
\begin{eqnarray}
\bar{v}_{\rm sw} & = &
\displaystyle{\frac{(P_4-P_1)^{1/2} (P_4-P_3)^{1/6}(2P_4+P_3)^{1/3}}{2^{10/3}\,3^{1/2}}} \frac{a}{\hbar}.\cr & & 
\label{velocity}
\end{eqnarray}
With the parameters given in Eq.~\ref{Results_microscopic_neutron_Analysis_4} we find $\bar{v}_{\rm sw}$ = 76\,(16)~m\,s$^{-1}$, in good agreement with the value ${v}_{\rm sw}$ = 84\,(2) and 82\,(2)~m\,s$^{-1}$ deduced in Sec.~\ref{Results_bulk_Cp_zero_field} from the analysis of the specific heat data.

Concerning the parameters derived from our diffuse scattering maps (Eq.~\ref{Results_microscopic_neutron_Analysis_3}), the magnon velocity cannot be computed since $(P_4 - P_1)<0$. At first sight, it might suggest that the parameters we infer from the two neutron maps are not reliable. However, the expression written in Eq.~\ref{velocity} is deduced from a linear spin-wave approximation. This approximation might not be reliable for a non-collinear magnet such as Er$_2$Ti$_2$O$_7$. This argument is based on theoretical results for triangular magnets.\cite{Zheng06a,Starykh06,Chernyshev06}

\section{Summary of our results and discussion} 
\label{Summary}

In this paper we have argued that the combined analysis of the values of the low-temperature Er$^{3+}$ magnetic moment and the two spectroscopic factors indicates that the moment can only be perpendicular (or close to perpendicular) to the local [111] axis. This analysis supports the magnetic structure proposed by Champion {\it et al.}\cite{Champion03}

The Er$_2$Ti$_2$O$_7$ magnetic susceptibility and specific heat have been carefully measured.
No magnetic hysteresis has been detected.  
The excitation gap is extremely small. The critical exponent of the specific heat in the
paramagnetic phase is typical for a three dimensional XY system. 
The specific heat data provides an estimate for the magnon velocity which
is in accord with a recently proposed model based on linear spin-wave theory.
Finally, concerning the bulk measurements, the magnetic phase diagram in the field-temperature plane has been determined up to 1.7~T for the three main crystal directions of a cube. 

The $\mu$SR data are consistent with the presence of a spin dynamics in the nanosecond time range in the ordered state. It might be associated with the short-range correlations detected by neutron diffraction in addition to the long-range order.\cite{Ruff08}

Diffuse neutron scattering data recorded in the paramagnetic state were analyzed in terms of a Hamiltonian accounting for all bilinear interactions between the spins in a tetrahedron. The four symmetry allowed interaction constants were determined and three of them were found 
to be positive, with $P_1$ and $P_4$ being the largest.
Note that the data analysis has assumed the  
interactions to be limited to the nearest neighbors Er$^{3+}$ ions. However, it must be recalled that interactions between further neighbors might be important. For instance they are determinant for the type of magnetic order adopted by Gd$_2$Ti$_2$O$_7$ relative to Gd$_2$Sn$_2$O$_7$ for which only nearest neighbors seem to matter.\cite{Wills06} The analysis of neutron scattering data for 
Dy$_2$Ti$_2$O$_7$ also confirms the importance of exchange interactions beyond nearest neighbors.\cite{Yavors08}

With the interaction constants and the spectroscopic factors known, it will be interesting to gauge
any proposed theoretical phase diagram to the fact that Er$_2$Ti$_2$O$_7$ does order magnetically at 
$T_{\rm N} = 1.23 \, (1)$~K. In addition, any reliable theory must be able to explain the short-range correlations
observed below $T_{\rm N}$ and their  nanosecond time scale dynamics. As it has been learned from
the study of Tb$_2$Sn$_2$O$_7$, neutron-spin echo measurements might be useful to further characterize
these exotic spin dynamics.\cite{Chapuis07,Rule09b}
The spin dynamics is the key feature which seems to distinguish a frustrated compound such as Er$_2$Ti$_2$O$_7$ from a conventional magnet.

\section*{Acknowledgments}

This research project has been partially supported by the European Science
Foundation through the Highly Frustrated Magnetism program. The $\mu$SR 
measurements were performed at S$\mu$S, Paul Scherrer 
Institute, Villigen, Switzerland, and at the ISIS facility, Rutherford Appleton Laboratory,
Chilton, UK. The neutron data were recorded at the Institut Laue Langevin, Grenoble, France.
We thank S.~S.~Sosin for his contribution to the specific heat measurements, J.A.~Hodges for ongoing discussions on Kramers doublets systems, M.E.~Zhitomirsky for numerous discussions on frustrated magnets, J.~Jerrett 
for technical assistance, and B. F{\aa}k for a careful reading of the manuscript. SC was supported by NSERC of Canada.

\appendix

\section{Physics of a tetrahedron of effective one-half spins 
embedded in a pyrochlore lattice}
\label{Tetrahedron}

Here we present a comprehensive quantum mechanical study 
of effective one-half spins embedded in a pyrochlore lattice.
After specifying the geometry, we describe the 
invariants which will enable us to build the Hamiltonian of the system. Then we explain the approximation that we use to compute the neutron diffuse scattering patterns. Finally, we compare our to 
others' equivalent Hamiltonian operators.

\subsection{Geometry}
\label{Tetrahedron_geometry}

For completeness, we first provide a description of a tetrahedron in a pyrochlore
lattice and of the spins at its corners. 

We choose a type A tetrahedron as the primitive unit cell; see Fig.~\ref{pyrochlore_structure}.
Each corner of the tetrahedron is occupied by a magnetic ion whose relative positions are given by
 ${\bf r}_1 = {a \over 4}(0,0,0)$, ${\bf r}_2 = {a \over 4}(1,1,0)$,
${\bf r}_3 = {a \over 4}(1,0,1)$ and ${\bf r}_4 = {a \over 4}(0,1,1)$, where $a$ is the
edge length of the cubic unit cell.
The distance between two magnetic ions is $d = a/(2\sqrt{2})$.
At each corner of the tetrahedron there is a local $D_{3d}$ symmetry, where the $C_3$ axis is one of the cube diagonals.

We denote the local $C_3$ axis at position $i$ as $z_i$ and 
define $x_i$ and $y_i$ axes to form a local orthogonal basis.
There is obviously some freedom in the choice of the $x_i$ and $y_i$ axes.
Below we list the unit vectors we have chosen for the bases at the four positions. For the first
two positions we have
\begin{eqnarray}
{\hat {\bf x}}_1  & = & (1,1,-2)/\sqrt{6}, \hspace{0.55 cm} {\hat {\bf x}}_2  = (-1,-1,-2)/\sqrt{6}, \cr
{\hat {\bf y}}_1  & = & (-1,1,0)/\sqrt{2}, \hspace{0.55 cm} {\hat {\bf y}}_2  = (1,-1,0)/\sqrt{2}, \cr
{\hat {\bf z}}_1  & = & (1,1,1)/\sqrt{3},  \hspace{0.85 cm} {\hat {\bf z}}_2  = (-1,-1,1)/\sqrt{3}, 
\label{Tetrahedron_geometry_2}
\end{eqnarray}
and at the remaining two positions we have
\begin{eqnarray}
{\hat {\bf x}}_3  & = & (-1,1,2)/\sqrt{6}, \hspace{0.40 cm} {\hat {\bf x}}_4  = (1,-1, 2)/\sqrt{6}, \cr
{\hat {\bf y}}_3  & = & (1,1,0)/\sqrt{2},  \hspace{0.70 cm} {\hat {\bf y}}_4  = (-1,-1,0)/\sqrt{2}, \cr
{\hat {\bf z}}_3  & = & (-1,1,-1)/\sqrt{3},\hspace{0.20 cm} {\hat {\bf z}}_4  = (1,-1,-1)/\sqrt{3}. 
\label{Tetrahedron_geometry_3}
\end{eqnarray}

A given spin ${\bf S}_i$ can be written using the cubic global axes as a basis or
the local axes at position $i$. We use upper case
superscripts to indicate components of the global basis, 
\begin{eqnarray}
{\bf S}_i = S^X_i{\hat {\bf X}} +  S^Y_i{\hat {\bf Y}}  +  S^Z_i{\hat {\bf Z}},
\label{Tetrahedron_geometry_4}
\end{eqnarray}
and lower case subscripts for components of a local frame,
\begin{eqnarray}
{\bf S}_i = S_{ix}{\hat {\bf x}}_i +  S_{iy}{\hat {\bf y}}_i  +  S_{iz}{\hat {\bf z}}_i.
\label{Tetrahedron_geometry_5}
\end{eqnarray}
Using the definitions above, we find
\begin{eqnarray}
S_{1x} & =& (S_1^X + S_1^Y - 2S_1^Z)/{\sqrt 6},\\
S_{1y} & = & (-S_1^X+S_1^Y)/{\sqrt 2}, \\
S_{1z} & = & (S_1^X+S_1^Y+S_1^Z)/{\sqrt 3},\\
S_{2x} & = & -(S_2^X +S_2^Y+2S_2^Z)/{\sqrt 6}, \\
S_{2y} & = & (S_2^X-S_2^Y)/{\sqrt 2}, \\
S_{2z} & = & (-S_2^X-S_2^Y+S_2^Z)/{\sqrt 3},\\
S_{3x} & =& (-S_3^X + S_3^Y + 2S_3^Z)/{\sqrt 6},\\ 
S_{3y} & = & (S_3^X+S_3^Y)/{\sqrt 2}, \\
S_{3z} & = & (-S_3^X+S_3^Y-S_3^Z)/{\sqrt 3},\\
S_{4x} & = & (S_4^X -S_4^Y+2S_2^Z)/{\sqrt 6}, \\
S_{4y} & = & (-S_4^X-S_4^Y)/{\sqrt 2}, \\
S_{4z} & = & (S_4^X-S_4^Y-S_4^Z)/{\sqrt 3},
\label{Tetrahedron_geometry_6}
\end{eqnarray}
and the inverse relations
\begin{eqnarray}
S_1^X &= & S_{1x}/{\sqrt 6} - S_{1y}/{\sqrt 2} + S_{1z}/{\sqrt 3},\\
S_1^Y &= & S_{1x}/{\sqrt 6} + S_{1y}/{\sqrt 2} + S_{1z}/{\sqrt 3},\\
S_1^Z &= & -2 S_{1x}/{\sqrt 6}  + S_{1z}/{\sqrt 3},\\
S_2^X &= & -S_{2x}/{\sqrt 6} + S_{2y}/{\sqrt 2} - S_{2z}/{\sqrt 3},\\
S_2^Y &= & -S_{2x}/{\sqrt 6} - S_{2y}/{\sqrt 2} - S_{2z}/{\sqrt 3},\\
S_2^Z &= & -2 S_{2x}/{\sqrt 6}  + S_{2z}/{\sqrt 3},\\
S_3^X &= & -S_{3x}/{\sqrt 6} + S_{3y}/{\sqrt 2} - S_{3z}/{\sqrt 3},\\
S_3^Y &= & S_{3x}/{\sqrt 6} + S_{3y}/{\sqrt 2} + S_{3z}/{\sqrt 3},\\
S_3^Z &= & 2 S_{3x}/{\sqrt 6}  - S_{3z}/{\sqrt 3},\\
S_4^X &= & S_{4x}/{\sqrt 6} - S_{4y}/{\sqrt 2} + S_{4z}/{\sqrt 3},\\
S_4^Y &= & -S_{4x}/{\sqrt 6} - S_{4y}/{\sqrt 2} - S_{4z}/{\sqrt 3}.\\
S_4^Z &= & 2 S_{4x}/{\sqrt 6}  - S_{4z}/{\sqrt 3}
\label{Tetrahedron_geometry_7}
\end{eqnarray}

\subsection{Determination of the invariants}
\label{Tetrahedron_invariants}
The interaction Hamiltonian ${\mathcal H}$ between the spins  
in the lattice is described by bilinear operators of 
the form $S_{i\alpha}S_{j\beta}$ where $i$ and $j$ are nearest neighbors.  
The space group $Fd\bar{3}m$ allows four different terms for a Kramers ion\cite{Curnoe07a,Curnoe08,McClarty2009} 
\begin{eqnarray}
{\mathcal H} =  
\tilde P_1 {\mathcal X}_1 + \tilde P_2 {\mathcal X}_2 + \tilde P_3 {\mathcal X}_3 + \tilde P_4 {\mathcal X}_4,
\label{Tetrahedron_invariants_1}
\end{eqnarray}
where $\tilde P_i$ are constants and
\begin{eqnarray}
{\mathcal X}_1 & = & -\frac{1}{3}\sum_{\langle i,j \rangle}S_{iz}S_{jz}, 
\label{Tetrahedron_invariants_2first}\\
{\mathcal X}_2 & = &  - \frac{\sqrt 2}{3}
\sum_{\langle i,j \rangle }\left [ \Lambda_{ij}(S_{iz}S_{j+}+S_{jz}S_{i+}) \right. \nonumber \\ 
& &\left.  + \Lambda_{ij}^{*}(S_{iz}S_{j-}+S_{jz}S_{i-}) \right ], \\
{\mathcal X}_3 & = &  \frac{1}{3}
\sum_{\langle i,j \rangle} \left (  \Lambda_{ij}^{*} S_{i+}S_{j+} + \Lambda_{ij}S_{i-}S_{j-} \right ), \\
{\mathcal X}_4 & = &  -\frac{1}{6}\sum_{\langle i,j \rangle} (S_{i+}S_{j-} + S_{j+}S_{i-}).
\label{Tetrahedron_invariants_2last}
\end{eqnarray}
The sum is over all the pairs of nearest neighbors (see 
Eq.~\ref{Results_bulk_sus_Weiss_3} for an example of the use of the notation in the case 
of the Heisenberg interaction).  
Below we give the expressions of the phase factors $\Lambda_{ij}$ introduced for 
${\mathcal X}_2$ and ${\mathcal X}_3$:
\begin{eqnarray}
\Lambda_{12} &=& \Lambda_{34} = 1,\\
\Lambda_{13} & = & \Lambda_{24} = \varepsilon = 
\exp \left( \frac{2\pi i}{3}\right) ,\\
\Lambda_{14} & = & \Lambda_{23} = \varepsilon^{*} = 
\exp \left( \frac{4\pi i}{3}\right).
\label{Tetrahedron_invariants_3}
\end{eqnarray}
Note that the sum of all four invariants is the isotropic exchange 
interaction $\sum_{i=1}^4{\mathcal X}_i = \sum_{\langle i,j \rangle} {\bf S}_i \cdot {\bf S}_j$ (obtained when $\tilde P_1 = \tilde P_2 = \tilde P_3 = \tilde P_4$).

To get some insight in the ${\mathcal H}$ expression, as an example, we  
consider the product ${\bf S}_1 \cdot {\bf S}_2$. Let us first focus on the 
contribution of the first invariant to  this product:
\begin{eqnarray}
{\mathcal X}_1 : \left(\frac{-1}{3}\right)\frac{1}{2}
\left ( S_{1z}S_{2z} + S_{2z}S_{1z}\right ) = -{1 \over 3} S_{1z}S_{2z}.
\nonumber
\label{Tetrahedron_invariants_4}
\end{eqnarray}
The contribution of the second invariant is easily found:
\begin{eqnarray}
{\mathcal X}_2 & :&  -2 {\sqrt{2} \over 3} \left ( S_{1z}S_{2x} + S_{2z}S_{1x} \right ).
\nonumber
\label{Tetrahedron_invariants_5}
\end{eqnarray}
In the same way the contributions of ${\mathcal X}_3$ and ${\mathcal X}_4$ 
can be derived. They are written in terms of the 
local Cartesian axes. Transforming to the global axes we derive
the expected relation
\begin{eqnarray}
{\bf S}_1 \cdot {\bf S}_2 =  S_1^X S_2^X +  S_1^Y S_2^Y + S_1^Z S_2^Z.
\nonumber
\label{Tetrahedron_invariants_6}
\end{eqnarray}

In its most general form (Eq.~\ref{Tetrahedron_invariants_1}) ${\mathcal H}$ includes symmetric (Heisenberg)
and the antisymmetric (Dzyaloshinskii-Moriya) interactions. 
For completeness, we give the expression of the Dzyaloshinskii-Moriya Hamiltonian.
In terms of the invariants, we derive 
\begin{eqnarray}
{\mathcal H}_{\rm DM} 
= -E_{\rm  t, DM} \left (4 {\mathcal X}_1 - \frac{1}{2}{\mathcal X}_2 +
{\mathcal X}_3 - 2{\mathcal X}_4 \right ),
\label{Tetrahedron_interactions_5}
\end{eqnarray}
where $E_{\rm  t, DM}$ scales the Dzyaloshinskii-Moriya interaction.

\subsection{Single tetrahedron approximation}
\label{ourmodel}
In order to compute
diffuse neutron scattering patterns exact eigenstates of ${\mathcal H}$ must be used,
however, the Hamiltonian given by Eq.\ \ref{Tetrahedron_invariants_1} is unsolvable in general.  
Therefore instead of using the full Hamiltonian (Eq.~\ref{Tetrahedron_invariants_1}), we
find exact solutions to an Hamiltonian ${\mathcal H}_{\rm t}$ restricted to tetrahedra of a single type.  This 
approach, which has been used to analyze diffuse neutron scattering patterns,\cite{Curnoe07a,Molavian07} is a considerable simplification of the original
model.  It amounts to replacing the sums in Eq.~\ref{Tetrahedron_invariants_1} over all tetrahedra, as it is implicit from the definition of the $\chi_i$'s, 
to a sum over all A-type (or all B-type) tetrahedra (Fig.~\ref{pyrochlore_structure}), as explained in 
Ref.~\onlinecite{Curnoe08}.  Thus each only three of the six nearest neighbors
of each Er atom are included in the calculation.  
Instead of the original interaction constants $\tilde P_i$
we introduce a similar set of constants $P_i$, such that
\begin{equation}
{\mathcal H}_{\rm t} = \sum_{\rm A \, tetrahedra} P_1 \mathcal{X}_1 + P_2 \mathcal{X}_2 + P_3 \mathcal{X}_3 + P_4 \mathcal{X}_4.
\label{singletet}
\end{equation}  
Here the sums appearing in the $\chi_i$ operators are limited to nearest neighbor spins belonging to {\em single} A tetrahedra, unlike in Eqs.~\ref{Tetrahedron_invariants_2first}-\ref{Tetrahedron_invariants_2last}.
While the full ramifications of this approximation are not understood,
at the very least we can estimate that the interaction constants $P_i$ 
are approximately a factor of two larger than $\tilde P_i$ to 
compensate for the missing exchange paths in Eq.\ \ref{singletet}.

\subsection{Relations between Hamiltonian parameters}
\label{Tetrahedron_parameters}

Other groups have proposed an Hamiltonian for the description of an effective one-half spin
system such as Er$_2$Ti$_2$O$_7$ for which the rare-earth crystal-field ground state is a Kramer's doublet. 

Ross {\em et al.}\cite{Ross11} and Savary {\em et al.} \cite{Savary12,Savary12a} write the Hamiltonian as
\begin{eqnarray}
{\mathcal H} & = & \sum_{\langle i,j \rangle} \left \{ J_{zz} S_{iz}S_{jz} -
J_{\pm} \left (S_{i+}S_{j-} + S_{i-}S_{j+} \right)  \right. \cr
&+ &  J_{\pm\pm} \left (\gamma_{ij} S_{i+}S_{j+} + \gamma^*_{ij}S_{i-}S_{j-} \right ) \cr
&+ & \left. J_{z\pm}  \left [S_{iz} \left (\zeta_{ij}S_{j+} + \zeta^*_{ij}S_{j-} \right ) 
 + i  \leftrightarrow  j  \right ] \right \},
\label{Tetrahedron_parameters_5}
\end{eqnarray}
where $\gamma$ and  $\zeta$ are $4 \times 4$ matrices:
\begin{eqnarray}
\zeta =
\begin{pmatrix}
   0          &  -1          & e^{i \pi/3}  & e^{-i \pi/3} \cr 
  -1          &  0           & e^{-i \pi/3} & e^{i \pi/3}  \cr
 e^{i \pi/3}   & e^{-i \pi/3} &  0           &  -1          \cr
 e^{-i \pi/3} & e^{i \pi/3}  &  -1          &  0           \cr
\end{pmatrix},  \, \, \gamma = - \zeta^* . \cr
\end{eqnarray}

Taking into account (i) the different labeling in the four sites of the tetrahedron as well as (ii) the different choice for the definition of the axes perpendicular to the local threefold axis which are adopted in these references compared to ours, the relations between the Hamiltonian parameters are
\begin{eqnarray}
J_{zz}\ & = & -{1 \over 3} \tilde P_1,  \hspace{0.55 cm}  J_{z\pm} =  {\sqrt{2} \over 3 } \tilde P_2 ,\cr
J_{\pm\pm} & = & {1 \over 3} \tilde P_3  , \hspace{0.85 cm} J_\pm = \  {1 \over 6} \tilde P_4 .
\label{Tetrahedron_parameters_6}
\end{eqnarray}
As explained in Sec.~\ref{ourmodel}, we expect $P_i \approx 2 \tilde P_i$.

Considering Yb$_2$Ti$_2$O$_7$ and others crystal-field ground-state doublet pyrochlore compounds, Onoda and Tanaka\cite{Onoda10,Onoda11a,Onoda11b} have also used an Hamiltonian equivalent to Eq.~\ref{Tetrahedron_parameters_5}, but with slight differences in the Hamiltonian parameters labeling.

Two other research groups  have studied the interaction of spins on the pyrochlore lattice with the purpose
to extract values of the interaction constants. However, since their interest was on the analysis 
of relatively high-temperature data, the full angular-momentum ${\bf J}_i$
was used.\cite{Malkin10,Thompson11a} Here we have recorded the diffuse scattering intensity at a 
sufficiently small temperature that it is justified to work within the effective one-half spin 
framework. It is possible to project the full angular momentum into the ground-state
Kramer's doublet. However, it seems that only semiformal relations between the parameters
of the effective  one-half spin and the full angular-momentum models can be derived.\cite{Ross11} 
Therefore we do not consider these relations in our analysis in
Sec.~\ref{Results_microscopic_neutron_Analysis}.

\section{Estimate of the electric field gradient acting on the $^{167}$Er nucleus}\label{Vzz}

Here we provide an estimate for the principal value of electric field gradient tensor $V_{zz}$. It is required for the computation of the nuclear contribution to the specific heat.

In an insulator $V_{zz}$ is written as the sum of two terms $V_{zz}^{\rm latt}$ and $V_{zz}^{4f}$, respectively, modeling the lattice and $4f$-shell contributions. The former contribution is expressed as $V_{zz}^{\rm latt}$ = $-\frac{4 A_2^0}{e}\frac{1-\gamma_\infty}{1-\sigma_2}$ where $A_2^0$ is a crystal-electric field parameter and $\gamma_\infty$ and $\sigma_2$ are a Sternheimer coefficient and the screening coefficient of the crystal field, respectively. From the literature values $A_2^0$ = 41.5\,(1.1)~meV\,$a_0^{-2}$ (Ref.~\onlinecite{Bertin12}) where $a_0$ = 52.92~pm is the Bohr radius and $(1-\gamma_\infty)/(1-\sigma_2)$ = 210\,(30) (Ref.~\onlinecite{Pelzl70}), we obtain $V_{zz}^{\rm latt}$ = $-1.24$\,(21)$\times 10^{22}$~V\,m$^{-2}$. The $4f$-shell contribution is written $V_{zz}^{4f}$ = $-\frac{e}{4\pi\varepsilon_0}\theta_2(1-R_Q)\langle r^{-3}\rangle_{4f}E_{4f}$. $R_Q$ is a Sternheimer coefficient, $\theta_2$ is a Stevens coefficient, and $\langle r^{-3}\rangle_{4f}$ and $E_{4f}$ are the expectation values respectively of the cube of the inverse distance between the nucleus and the Er$^{3+}$ $4f$ shell, and the quadrupole operator $3J_z^2-J(J+1)$ acting on the $4f$ shell. As usual, $\varepsilon_0$ is the permittivity of free space. From the literature values $R_Q$ = 0.29\,(1) (Ref.~\onlinecite{Pelzl70}), $\theta_2$ = $2.54\times 10^{-3}$, $\langle r^{-3}\rangle_{4f}$ = $11.36\,a_0^{-3}$ (Ref.~\onlinecite{Freeman79}) and $E_{4f}$ = $-7.86$ (Ref.~\onlinecite{Bertin12}), we get $V_{zz}^{4f}$ = $1.6\times 10^{21}$~V\,m$^{-2}$. 
Summing up the two contributions we obtain $V_{zz}$ = $-1.1\,(3)\times 10^{22}$~V\,m$^{-2}$.

\bibliography{reference.bib}

\end{document}